\documentclass[12pt]{article}
\pdfoutput=1 
\usepackage{jheppub}
\usepackage{iip-arxiv}
\usepackage[utf8]{inputenc}
\usepackage{bm}
\usepackage{ifthen}
\usepackage{breqn}

\usepackage{array}

\DeclareMathAlphabet{\mathbbold}{U}{bbold}{m}{n}
 \DeclareMathOperator{\Tr}{Tr}
\DeclareMathOperator*{\res}{\mathrm{Res}}
\DeclareMathOperator{\sgn}{sgn}
\newcommand{\slc}{\mathrm{SL}(2,\mathbb{C})}

\newcommand{\del}[1][z]{\partial_{#1}}
\newcommand{\Dfrac}[3][]{\frac{d^{#1}#2}{d{#3}^{#1}}}
\newcommand{\Dpfrac}[3][]{\frac{\partial^{#1} #2}{\partial {#3}^{#1}}}
\newcommand{\cblockc}[1][\bm
P]{\mathcal{F}_{#1}(\Delta_{L},\Delta_{H},\bm{\Delta}|z,\lambda,\bm{z})}
 \newcommand{\cblock}[1][\bm P]{\mathcal{F}_{#1}}
\newcommand\pfraction[4]{\ifthenelse{\equal{#1}{}}{\frac{#4}{#2-#3}}{\frac{#4}{(#2-#3)^{#1}}}}
\newcommand{\lmtheta}[2]{\,\theta_0,\theta_1,\theta_t\ifthenelse{\equal{#1}{}}{}{+#1},\theta_\infty\ifthenelse{\equal{#2}{}}{}{+#2}}
\newcommand*{\defeq}{\mathrel{\vcenter{\baselineskip0.5ex
      \lineskiplimit0pt \hbox{\scriptsize.}\hbox{\scriptsize.}}}%
  =} \newcommand{\btheta}{\bm{\theta}}
\newcommand{\cbar}{\bar{C}_{n}(\btheta,\sigma)}

\title{Classical Conformal Blocks and Accessory Parameters from Isomonodromic Deformations}

\author[1]{Máté Lencsés} 
\author[2]{and Fábio Novaes}
\affiliation{International Institute of Physics, Federal University of Rio
Grande do Norte, Campus Universitário, Lagoa Nova, Natal-RN 59078-970, Brazil }
\email{matelencses at gmail.com}
\email{fabionsantos at gmail.com}

\abstract{Classical conformal blocks appear in the large central
  charge limit of 2D Virasoro conformal blocks. In the
  $AdS_{3}/CFT_{2}$ correspondence, they are related to classical bulk
  actions and used to calculate entanglement entropy and geodesic
  lengths. In this work, we discuss the identification of classical
  conformal blocks and the Painlevé VI action showing how
  isomonodromic deformations naturally appear in this context.  We
  recover the accessory parameter expansion of Heun's equation from
  the isomonodromic $\tau$-function. We also discuss how the $c=1$
  expansion of the $\tau$-function leads to a novel approach to
  calculate the 4-point classical conformal block.}

\preprint{}

\begin{document}

\maketitle

\section{Introduction}\label{sec:intro}

Classical conformal blocks
\cite{Zamolodchikov1987,Zamolodchikov1996,Litvinov:2013sxa} are
essential pieces of the holographic duality between $AdS_{3}$ gravity
and 2D conformal field theory (CFT)
\cite{Brown:1986nw,Hartman2013,Fitzpatrick2014}. Holographic CFTs are
assumed to have a \emph{sparse light spectrum} and to contain only
correlators dominated by the identity channel in the semiclassical
limit \cite{Hartman2013,Hartman:2014oaa}, usually called classical
vacuum blocks. These blocks can be used to study thermal aspects of 3D
gravity \cite{Fitzpatrick2014,Fitzpatrick2015a}, holographic
entanglement entropy \cite{Ryu:2006bv,Hartman2013}, to calculate bulk
geodesic lengths
\cite{Hijano2015,Hijano2015a,Alkalaev:2015lca,Alkalaev:2016rjl,Chen:2016dfb}
and Lyapunov exponents of out-of-time-order correlators
\cite{Roberts:2014ifa}.

Classical conformal blocks are defined by the conformal block
exponentiation conjecture in the large central charge
limit\footnote{There is no rigorous CFT proof of this conjecture,
  only plausibility and numerical arguments
  \cite{Zamolodchikov1987}. It is also compatible with classical
  saddle-point arguments in Liouville theory \cite{Harlow:2011ny,Dong:2018esp}.}
\cite{Zamolodchikov1987,Harlow:2011ny,Fitzpatrick2014}
\begin{equation}
  \label{eq:intro1}
  \langle V_{\Delta_{0}}(0)V_{\Delta_{x}}(x)\Pi_{\Delta} V_{\Delta_{1}}(1)V_{\Delta_{\infty}}(\infty) \rangle_{c\rightarrow \infty} \sim \exp{
    \left(
      \frac{c}{6}f_{\delta}(\delta_{0}, \delta_{x}, \delta_{1}, \delta_{\infty};x)
    \right)}.
\end{equation}
We denote the central charge by $c$ and
$\Delta_{i}= \frac{c}{6} \delta_{i}$, $i=0,x,1,\infty$, are the
conformal dimensions of the chiral primary operators $V_{\Delta_{i}}$,
with $\delta_{i}$ being the classical dimensions. $\Pi_{\Delta}$ is
the projection operator to the intermediate channel with weight
$\Delta = \frac{c}{6} \delta$. In broad terms, we say that an operator
$\mathcal{O}$ is \emph{light} if its weight
$\Delta_{\mathcal{O}} \ll c$ and it is called \emph{heavy} if
$\Delta_{\mathcal{O}} \sim c$ as $c \rightarrow \infty$. The function
$f_{\delta}(\{\delta_{i}\};x)$ is called the \emph{classical conformal
  block} \cite{Zamolodchikov1996,Litvinov:2013sxa}. No closed CFT
expression is known for this special function. It can be written as a
series expansion in $x$ using a direct CFT approach \cite{Belavin1984}
or Zamolodchikov's recurrence formula
\cite{Zamolodchikov1984,Zamolodchikov1987}. The CFT approach quickly
gets too cumbersome to find the explicit coefficients at higher orders
in $x$. A formal resummation of the recurrence formula was presented
in \cite{Perlmutter2015}, but the full classical conformal block is
still out of reach.

A more promising direction is to obtain classical conformal blocks via
the AGT correspondence \cite{Alday2010b}. The Nekrasov partition
function $Z_{\text{Nek}}$ 
encodes information about the moduli space of vacua and its
non-perturbative corrections in supersymmetric gauge
theories \cite{Nekrasov2003a}.  According to the AGT correspondence,
for a certain class of $\mathcal{N}=2$ SUSY theories, $Z_{\text{Nek}}$
is identified with a 2D Liouville conformal block. One can show that
the so-called Nekrasov-Shatashvili limit \cite{Nekrasov2009}
is equivalent to the large central charge limit of Liouville theory.
This fact was used in \cite{Piatek2011,Ferrari2012a} to obtain an
expression for the classical conformal block in terms of the
$\mathcal{N}=2$ \emph{twisted superpotential}, calculated at the
saddle-point of the partition function. The saddle-point condition
then has to be solved order-by-order in $x$.
The twisted superpotential can be understood in terms of the symplectic
geometry of the moduli space of SL$(2,\mathbb{C})$ flat-connections and the
Bethe/gauge correspondence \cite{Nekrasov:2011bc}, which gives extra
hints on the deeper mathematical structure of classical conformal
blocks. For a review on exact results in $\mathcal{N}=2$ field
theories, see \cite{Teschner2016}.

In a parallel development, Litvinov \emph{et al}
\cite{Litvinov:2013sxa} discussed how the 4-point classical conformal
block is related to the classical action of the Painlevé VI (PVI)
equation \cite{Guzzetti2012b}.  In this approach, the derivative of
the PVI action, evaluated on a PVI solution with certain boundary
conditions, gives the accessory parameter of a Fuchsian differential
equation with 4 regular singular points, also known as Heun's equation
\cite{ronveaux1995heun}. 
The Heun equation is obtained by the classical limit of the 5-point
conformal block with a level-2 \emph{light} degenerate insertion
\begin{equation}
  \label{eq:intro2}
  \langle \varphi_{L}(z)V_{\Delta_{0}}(0)V_{\Delta_{x}}(x)\Pi_{\Delta} V_{\Delta_{1}}(1)V_{\Delta_{\infty}}(\infty)\rangle_{c\rightarrow \infty} 
  \sim \psi(z,x) \exp{
    \left(
      \frac{c}{6}f_{\delta}(\{\delta_{i}\};x)
    \right)}.
\end{equation}
The level-2 null vector equation for $\varphi_{L}(z)$, also known as
level-2 BPZ equation \cite{Belavin1984}, reduces to the normal form of
Heun's equation
\begin{multline}
  \label{eq:heun}
  \left[ \del[z]^{2} - \frac{t(t-1)H_{x}}{z(z-1)(z-x)}  
    +\frac{\delta_{0}}{z^{2}}+\pfraction{2}{z}{1}{\delta_{1}}+\pfraction{2}{z}{x}{\delta_{x}}+\frac{\delta_{\infty}-\delta_{0}-\delta_{1}-\delta_{x}}{z(z-1)}
\right]
  \psi(z,x) = 0,
\end{multline}
where the accessory parameter is given by
\begin{equation}
  \label{eq:heunaccessory}
  H_{x} = -\del[x]f_{\delta}(\{\delta_{i}\};x).
\end{equation} 
On the other hand, the semiclassical limit of the 5-point conformal
block with a level-2 heavy degenerate insertion
\begin{equation}
  \label{eq:intro3}
  \langle \varphi_{H}(\lambda)V_{\Delta_{0}}(0)V_{\Delta_{x}}(x)\Pi_{\Delta} V_{\Delta_{1}}(1)V_{\Delta_{\infty}}(\infty) \rangle_{c\rightarrow \infty} 
  \sim  \exp{
    \left(
      \frac{c}{6}S_{\delta}(\{\delta_{i}\};\lambda,x)
    \right)}
\end{equation}
obeys a BPZ equation equivalent to the Hamilton--Jacobi equation of
the PVI action.  This means that the BPZ equation implies that
$\lambda$ must be a solution of the PVI equation. 

In order to recover the 4-point classical conformal block from the PVI
action, the authors of \cite{Litvinov:2013sxa} set $\lambda=\infty$ in
\eqref{eq:intro3} and obtain an integral formula for the classical
block. Taking the derivative of this formula then leads to the
accessory parameter of Heun's equation \eqref{eq:heunaccessory} in
terms of the initial condition for $\lambda$, fixed by
$\lambda=\infty$. With a clever usage of a double series expansion of the
PVI solution \cite{Guzzetti2012b}, the authors of
\cite{Litvinov:2013sxa} managed to solve the condition
$\lambda=\infty$ order by order in $x$ and then substituted the result
into the accessory parameter formula. However, their procedure relies
on substituting the double series expansion into the PVI equation,
which is a complicated second order non-linear differential equation,
to obtain the doubles series expansion terms also order by order in
$x$.

In this paper, we show how the isomonodromic $\tau$-function
\cite{Jimbo1981b,Jimbo:1981-2,Jimbo:1981-3}, also known as PVI
$\tau$-function~\cite{Jimbo:1982,Gamayun2012}, can be used to find the
accessory parameter expansion discussed in \cite{Litvinov:2013sxa} in
a more straightforward way. Both PVI solutions and the accessory
parameter can be written in terms of the $\tau$-function. 
Since the works of Sato, Jimbo and Miwa \cite{Sato1978, Sato1979,
  Sato1979b, Sato1979a, Sato1980}, it is known that the PVI
$\tau$-function is related to a $c=1$ correlator of monodromy
fields. In fact, the isomonodromic approach effectively solves the
Riemann-Hilbert problem of $\slc$ Fuchsian systems
\cite{Jimbo1981b,Jimbo:1981-2,Jimbo:1981-3}. For a long time, only the
asymptotics of the $\tau$-function was known, limiting its scope of
applications.  Its full expansion was constructed only recently in
\cite{Gamayun2012} and proved in \cite{Iorgov2015}. This expansion is
given by a linear combination of $c=1$ conformal blocks, which are
written in closed form via AGT correspondence. There are two
integration constants $(\sigma,s)$ for the $\tau$-function, labeling
irreducible representations of the 4-point monodromy group, which we
review in section \ref{sec:moduli-space-flat}. The relevant expansion
in terms of $c=1$ conformal blocks \cite{Gamayun2012} is presented in
\ref{sec:isom-equat-c=1}. This $\tau$-function expansion is what
allows us to solve the initial condition for the isomonodromic flow
and then find the accessory parameter expansion.

The way we solve the accessory parameter expansion clarifies two
important things. First, the initial condition used by
\cite{Litvinov:2013sxa} can be uniquely defined in terms of
isomonodromic deformations. Second, the recent result
\cite{Gamayun2012} on the $c=1$ expansion of the PVI $\tau$-function
allow us to present the accessory parameter expansion in a more
systematic way, providing a practical algorithm on how to fully solve
this problem.

Let us now give the outline of this paper. In section
\ref{sec:isom-semicl-limit}, we review the relationship between the
semiclassical limit of BPZ equations and Fuchsian equations. We
introduce a slightly more general derivation than
\cite{Litvinov:2013sxa}, using a 6-point conformal block with 2
degenerate insertions instead of a 5-point conformal block. This
introduces the relationship between the classical conformal block
\eqref{eq:intro3} and the Painlevé VI action.

In Section \ref{sec:isom-tau-funct}, we review the standard setup of
isomonodromic deformations, the $\tau$-function definition and the
associated ordinary differential equation (ODE) with one extra
apparent singularity. We then review the connection between the
semiclassical limit of CFT correlators and isomonodromic deformations
\cite{Teschner2011,Teschner2017a}. We move on to discuss the
relationship between Fuchsian equations and the monodromy group of the
4-punctured sphere, summarizing how the different parameters in this
paper relate with each other. Finally, we show how the BPZ equations
of section \ref{sec:isom-semicl-limit} can be encoded in a Fuchsian
system when $c=1$.  The conclusion is that the monodromy data of a
$c=\infty$ conformal block can be encoded in a $c=1$ Fuchsian
system. This makes the connection between classical conformal blocks,
Painlevé VI and isomonodromic deformations explicit.

In section \ref{sec:access-param-from}, we present our algorithm to
calculate the accessory parameter \eqref{eq:heunaccessory} using the
isomonodromic $\tau$-function \cite{Jimbo:1982,Gamayun2012}.  The
algorithm consists of three steps, described in detail in this
section. The crucial step is to solve a special initial condition to
the isomonodromic flow. We can only solve this condition
order-by-order in $x$, in a similar fashion to \cite{Litvinov:2013sxa}
or the AGT approach of \cite{Ferrari2012a}. This constrains the
general moduli space of the monodromy group to a subspace with only
one parameter, the composite monodromy $\sigma$. Our approach gives
new analytic insights on classical conformal blocks compared to
\cite{Litvinov:2013sxa}.

In Section \ref{sec:corr-from-access}, we show how the PVI action can
be written only in terms of $\tau$-functions.  In principle, this
gives a formula for the 4-point classical conformal block, up to the
solution of the appropriate PVI initial condition. However, as we discuss in the linear
dilaton case, the PVI $\tau$-function simplifies in certain special
cases \cite{Gamayun2012,Gamayun:2013auu}.
We leave the study of the different special cases for
future work.

The relationship between isomonodromic deformations and the CFT
semiclassical limit has been previously explored in
\cite{Teschner2011,Teschner2017a}. Although these papers recognize the
importance of isomonodromic deformations, the relevance of the
$\tau$-function and its detailed implementation has only been discussed
here. Therefore, our main contribution is to give a unified
prescription on how to use the isomonodromic $\tau$-function to obtain
the accessory parameter of Heun's equation and the associated
classical conformal block in 2D CFT.

\section{Classical Conformal Blocks and Accessory Parameters}
\label{sec:isom-semicl-limit}

In this section, we review how the semiclassical limit of a special
6-point conformal block leads to a Fuchsian equation with 4 singular
points and one apparent singularity. The method generalizes to
arbitrary $n$-point conformal blocks with an appropriate number of
extra degenerate insertions \cite{Teschner2011,Teschner2017a}.

Our derivation relies only on the 2D conformal symmetry and the
definition of conformal blocks.  Although we use the standard
Liouville notation for the conformal dimensions and central charge, we
do not make any particular assumption about the spectrum. This comes
\emph{a posteriori} and it is the main point of the bootstrap program
\cite{Zamolodchikov1996,Hartman:2014oaa,Fitzpatrick2014,Chang2016,Chang2016a,Collier2016,Collier2017}. For
reviews on Liouville theory and CFT, we suggest
\cite{Ribault:2014aa,Teschner2017}.

A chiral primary operator $V_{\Delta(P)}$ has conformal dimension
\begin{equation}
  \label{eq:1}
  \Delta(P) = \frac{Q^{2}}{4} + P^{2},
\end{equation}
where $P$ is the momentum of the operator and $Q$ parametrizes the
central charge as
\begin{equation}
  \label{eq:2}
  c = 1 + 6Q^{2},\quad Q = b + \frac{1}{b},
\end{equation}
with $b\in \mathbb{C}$. The spectrum is dual under
$b \rightarrow 1/b$,
and we choose $b\rightarrow 0$ to denote the semiclassical
limit.

\begin{figure}[t]
  \centering
  \includegraphics[width=.8\textwidth]{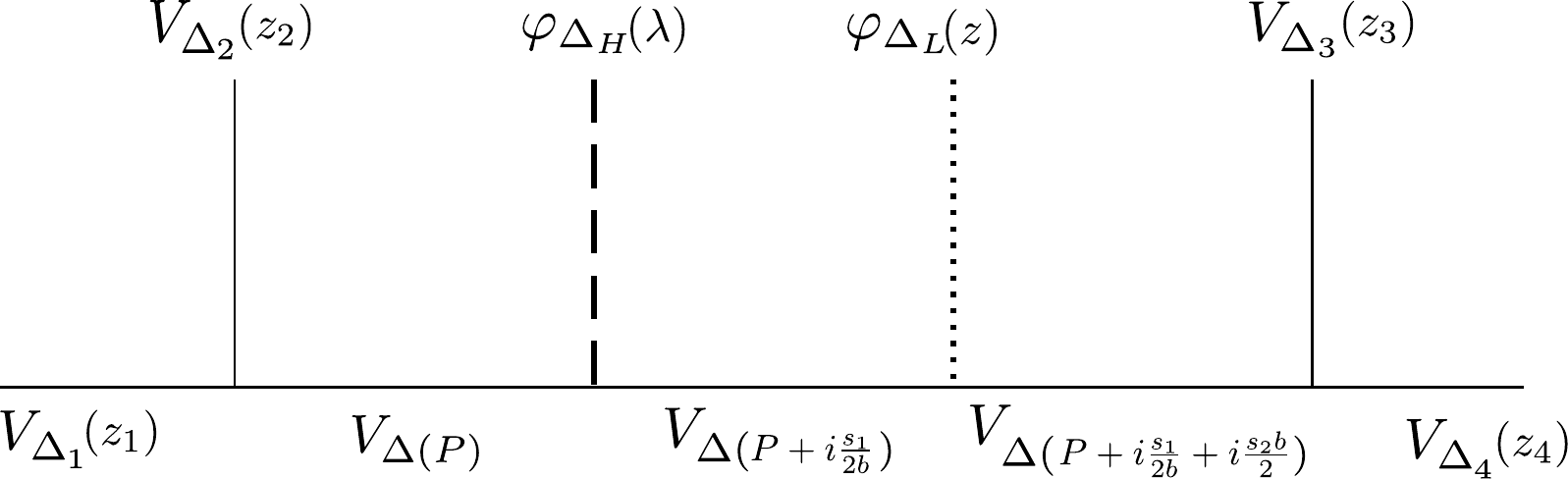}
  \caption{6-point conformal block with one heavy and one light
    insertion. The intermediate momenta are labeled by $P$ and two
    integers $s_{1},s_{2}=\pm 1$. }
  \label{fig:6ptblock}
\end{figure}

Let us consider the chiral
6-point correlator with two degenerate insertions
\begin{equation}
  \label{eq:6ptcorrelator}
  \langle
  \varphi_{\Delta_{L}}(z)\varphi_{\Delta_{H}}(\lambda)V_{\Delta_{1}}(z_{1})V_{\Delta_{2}}(z_{2})V_{\Delta_{3}}(z_{3})V_{\Delta_{4}}(z_{4})
  \rangle
  =\sum_{\bm P}\mathcal{C}_{\bm P}\,\cblockc,
\end{equation}
where $\varphi_{\Delta_{L}}$ and $\varphi_{\Delta_{H}}$ stands for
\emph{light} and \emph{heavy} level-2 degenerate operators,
respectively. 
$\bm\Delta = (\Delta_{1},\ldots,\Delta_{4})$ stands for the conformal
dimensions of generic heavy fields at positions
$\bm{z} = (z_{1},\cdots,z_{4})$. The conformal dimensions can be
written as
\begin{align}
  \label{eq:weights}
 \Delta_{L} = -\frac{1}{2} -\frac{3b^{2}}{4},\quad \Delta_{H} = -\frac{1}{2} -\frac{3}{4b^{2}},\quad
  \Delta_{i} \equiv \Delta(P_{i}) = \frac{Q^{2}}{4}+P_{i}^{2},\ (i=1,\ldots, 4).
  \end{align}
  $\mathcal{C}_{\bm{P}}$ represents the appropriate products of
  structure constants. With the ordering given
  in figure~\ref{fig:6ptblock}, the conformal block $\cblockc$ is labeled by
  the intermediate momenta
  $\bm P = (P,P+\tfrac{i s_{1}}{2b},P+\tfrac{i s_{1}}{2b}+\tfrac{i
    s_{2}b}{2})$, $s_{1},s_{2} = \pm 1$.

The correlator \eqref{eq:6ptcorrelator} obeys a light BPZ equation in
the variable $z$ and a heavy BPZ equation in the variable
$\lambda$. Because of linearity, the conformal blocks obey the same
equations
\begin{subequations}
\label{eq:BPZ}
  \begin{align}
  \left[
  \frac{1}{b^{2}}\del^{2} +
  \pfraction{2}{z}{\lambda}{\Delta_{H}}+
  \pfraction{}{z}{\lambda}{\del[\lambda]}+\sum_{i=1}^{4}
  \left(
  \pfraction{2}{z}{z_{i}}{\Delta_{i}}+\pfraction{}{z}{z_{i}}{\del[z_{i}]}
  \right)
 \right]\cblock(z,\lambda,\bm{z})&=0, \label{eq:BPZa}\\[10pt]
   \left[b^{2}\del[\lambda]^{2} +
  \pfraction{2}{\lambda}{z}{\Delta_{L}}+
  \pfraction{}{\lambda}{z}{\del[z]}+\sum_{i=1}^{4}
  \left(
  \pfraction{2}{\lambda}{z_{i}}{\Delta_{i}}+\pfraction{}{\lambda}{z_{i}}{\del[z_{i}]}
  \right)
 \right]\cblock(z,\lambda,\bm{z}) \label{eq:BPZb}&=0,
\end{align}
\end{subequations}
where we omitted the conformal dimensions in
$\cblock(z,\lambda,\bm{z}) $ for convenience. We can simplify these
equations using global conformal transformations and the Ward identity
\begin{multline}
  \label{eq:wardid}
  \langle T(w)
  \varphi_{L}(z)\varphi_{H}(\lambda)V_{\Delta_{1}}(z_{1})\cdots V_{\Delta_{4}}(z_{4})\rangle=\\[5pt]
  = \sum_{i=L,H,1,\dots,4} \left(
    \pfraction{2}{w}{z_{i}}{\Delta_{i}}+\pfraction{}{w}{z_{i}}{\del[z_{i}]}
  \right) \langle \varphi_{L}(z)\varphi_{H}(\lambda)V_{\Delta_{1}}(z_{1})\cdots
  V_{\Delta_{4}}(z_{4})\rangle,
\end{multline}
where $T(w)$ is the 2D stress tensor, with $z_{L}=z$ and
$z_{H}=\lambda$.  A straightforward consequence of the asymptotic
behavior $T(w) \sim w^{-4}$ as $w\rightarrow \infty$ is that
\begin{equation}
  \label{eq:6}
  \oint_{w=\infty} dw\; \epsilon(w) \langle T(w)
  \varphi_{L}(z)\varphi_{H}(\lambda)V_{1}(z_{1})\cdots V_{4}(z_{4})\rangle = 0
\end{equation}
with $\epsilon(w) = \prod_{i=1}^{3}(w-z_{i})/(w-z)$
\cite{Ribault:2014aa}.  Using \eqref{eq:wardid} and \eqref{eq:6} in
\eqref{eq:BPZa}, we choose $z_{1}=0, z_{2}=t, z_{3}=1, z_{4}=\infty$
and relabeling the $\Delta$'s accordingly, we get
\begin{multline}
  \label{eq:simpleBPZlight}
  \left[ b^{-2}\del[z]^{2}- \left( \frac{1}{z}+\frac{1}{z-1}
    \right)\del[z] +
    \frac{\Delta_{0}}{z^{2}}+\pfraction{2}{z}{1}{\Delta_{1}}+\pfraction{2}{z}{t}{\Delta_{t}}+\pfraction{2}{z}{\lambda}{\Delta_{H}} + \frac{t(t-1)\del[t]}{z(z-1)(z-t)}+\right.\\[5pt]
  \left. +
    \frac{\lambda(\lambda-1)\del[\lambda]}{z(z-1)(z-\lambda)}+\frac{\Delta_{\infty}-\Delta_{L}-\Delta_{H}-\Delta_{0}-\Delta_{1}-\Delta_{t}}{z(z-1)}
\right]
  \cblock(z,\lambda,t) = 0.
\end{multline}

Let us now analyze the semiclassical limit. Assuming $P_{k} = i \theta_{k}/b$ and $P= i \sigma/b$ as $b\rightarrow 0$, we have
\begin{equation}
  \label{eq:weightssclimit}
  \Delta_{L} \rightarrow -\frac{1}{2},\quad
  \Delta_{H} \rightarrow -\frac{3}{4b^{2}},\quad  \Delta(P) \rightarrow \frac{\delta_{\sigma}}{b^{2}},\quad
  \Delta_{k} \rightarrow \frac{\delta_{k}}{b^{2}},\quad ( k= 0,x,1,\infty),
\end{equation}
where
\begin{equation}
  \label{eq:25}
  \delta_{\sigma} = \frac{1}{4}-\sigma^{2},\quad\delta_{k} =
  \frac{1}{4}-\theta_{k}^{2},\quad (k= 0,x,1,\infty).
\end{equation}
Assuming heavy-light factorization and exponentiation\footnote{The
  exponentiation can be understood in Liouville field theory. In
  principle, the semiclassical limit of a correlator is given by a
  classical saddle-point of the Liouville action, if this saddle-point
  is unique.}
\cite{Zamolodchikov1996,Harlow:2011ny,Litvinov:2013sxa}, we write the
semiclassical limit of the 6-point conformal block when
$b\rightarrow 0$ as
\begin{equation}
  \label{eq:sclimit6pt}
  \cblock[\sigma]^{\pm}(z,\lambda,t) \sim \psi(z , \lambda,t)\exp{
    \left(
      \frac{1}{b^{2}}\,S_{\sigma}^{\pm}(\lambda,t)
    \right)}.
\end{equation}
We simplified the notation to $\bm{P} \rightarrow (\sigma,\pm)$
because the light field $\varphi_{L}(z)$ does not contribute to the
intermediate momenta in the semiclassical limit.  The function
$\psi(z,\lambda,t)$ encodes the fusion (monodromy) information of the
light field, as we are going to see below.  Fusing the light
degenerate field with any of the other fields,
we end up with the semiclassical limit of the 5-point
block with all insertions being heavy
\begin{equation}
  \label{eq:sclimit5pt}
  \cblock[\sigma]^{\pm}(\lambda,t) \sim 
  \exp{
    \left(
      \frac{1}{b^{2}}\,S_{\sigma}^{\pm}(\lambda,t)
    \right)}.
\end{equation}
Substituting \eqref{eq:sclimit6pt} in \eqref{eq:simpleBPZlight}
gives
\begin{multline}
  \label{eq:sclimitBPZlight}
  \left[ \del[z]^{2} + \frac{t(t-1)C_{t}}{z(z-1)(z-t)} +
    \frac{\lambda(\lambda-1)C_{\lambda}}{z(z-1)(z-\lambda)}+ \right.\\[5pt]
  \left. +
    \frac{\delta_{0}}{z^{2}}+\pfraction{2}{z}{1}{\delta_{1}}+\pfraction{2}{z}{t}{\delta_{t}}-\pfraction{2}{z}{\lambda}{\tfrac34}+\frac{\delta_{\infty}-\delta_{0}-\delta_{1}-\delta_{t}+\tfrac34}{z(z-1)}
\right]
  \psi(z,\lambda,t) = 0,
\end{multline}
which is a Fuchsian equation with 5 singular points and
\emph{accessory parameters}
\begin{equation}
  \label{eq:9}
 C_{t} = \del[t]S_{\sigma}^{\pm},\quad C_{\lambda} = \del[\lambda]S_{\sigma}^{\pm}.
\end{equation}

Following the same procedure for \eqref{eq:BPZb}, we get
\begin{multline}
  \label{eq:simpleBPZheavy}
  \left[ b^{2}\del[\lambda]^{2}- \left(
      \frac{1}{\lambda}+\frac{1}{\lambda-1} \right)\del[\lambda] +
    \frac{\Delta_{0}}{\lambda^{2}}+\pfraction{2}{\lambda}{1}{\Delta_{1}}+\pfraction{2}{\lambda}{t}{\Delta_{t}}+\pfraction{2}{\lambda}{z}{\Delta_{L}}+  \frac{t(t-1)\del[t]}{\lambda(\lambda-1)(\lambda-t)} +\right.\\[5pt]
  \left. + \frac{z(z-1)\del[z]}{\lambda(\lambda-1)(\lambda-z)}
    +\frac{\Delta_{\infty}-\Delta_{L}-\Delta_{H}-\Delta_{0}-\Delta_{1}-\Delta_{t}}{\lambda(\lambda-1)}
\right]
  \cblock[\bm{P}](z,\lambda,t) = 0.
\end{multline}
The semiclassical limit of this equation gives a constraint on the
accessory parameters \eqref{eq:9}
\begin{multline}
  \label{eq:sclimitBPZheavy}
  (\del[\lambda]S_{\sigma}^{\pm})^{2}- \left(
    \frac{1}{\lambda}+\frac{1}{\lambda-1}
  \right)\del[\lambda]S_{\sigma}^{\pm} +
  \frac{t(t-1)\del[t]S_{\sigma}^{\pm}}{\lambda(\lambda-1)(\lambda-t)} +
  \\[5pt]
  +
  \frac{\delta_{0}}{\lambda^{2}}+\pfraction{2}{\lambda}{1}{\delta_{1}}+\pfraction{2}{\lambda}{t}{\delta_{t}}+\frac{\delta_{\infty}-\delta_{0}-\delta_{1}-\delta_{t}+\tfrac34}{\lambda(\lambda-1)}
  = 0.
\end{multline}
This is exactly the condition for $z=\lambda$ to be an apparent
singularity of \eqref{eq:sclimitBPZlight}
\cite{Iwasaki:1991,Teschner2011,Teschner2017a}.  This means that
$\psi(z,\lambda,t)$ has integer monodromy around $z=\lambda$ but no
logarithmic behavior. Thus $z=\lambda$ is \emph{not} a singular
point of the solution.  Moreover, \eqref{eq:sclimitBPZheavy} can be
interpreted as a Hamilton--Jacobi equation for
$S_{\sigma}^{\pm}(\lambda,t)$
\begin{equation}
  \label{eq:taufunctioneq}
  \Dpfrac{S_{\sigma}^{\pm}}{t} + H
  \left(\lambda,\Dpfrac{S_{\sigma}^{\pm}}{\lambda},t \right) = 0,
\end{equation}
where
\begin{multline}
\label{eq:hamiltonian}
H(\lambda,p,t) =\frac{\lambda(\lambda-1)(\lambda-t)}{t(t-1)} \left[
  p^{2}- \left( \frac{1}{\lambda}+\frac{1}{\lambda-1} \right)p
  +\right.
\\[5pt]
\left.  +
  \frac{\delta_{0}}{\lambda^{2}}+\pfraction{2}{\lambda}{1}{\delta_{1}}+\pfraction{2}{\lambda}{t}{\delta_{t}}+\frac{\delta_{\infty}-\delta_{0}-\delta_{1}-\delta_{t}+\tfrac34}{\lambda(\lambda-1)}
\right].
\end{multline}
The equation of motion for $\lambda(t)$ obtained from this Hamiltonian
is the PVI equation, therefore $S_{\sigma}^{\pm}(\lambda,t)$ is the PVI
action. The heavy BPZ equation 
gives a saddle-point
condition for $S_{\sigma}^{\pm}(\lambda,t)$ \cite{Litvinov:2013sxa}.

If we define the Hamiltonian system $(\lambda(t),p(t))$ evolving under
the PVI Hamiltonian \eqref{eq:hamiltonian}, it is possible to show
that the monodromy data of the Fuchsian equation
\eqref{eq:sclimitBPZlight} \emph{does not change} as we change $t$ in
the complex plane.  Therefore, \eqref{eq:sclimitBPZlight} is the
\emph{isomonodromic deformation} of a 4-point Fuchsian equation, a
deformed Heun's equation, with $z=\lambda$ being an apparent
singularity and not contributing to the monodromy
data~\cite{Iwasaki:1991,Teschner2011,Teschner2017a}. This means that
isomonodromic deformations naturally emerge in CFT. We will review the
standard setup of isomonodromic deformations in the next section. We
also discuss how isomonodromic deformations relate the monodromy group
of the 4-punctured sphere and the moduli space of Fuchsian equations
(also called \emph{opers} in the literature \cite{Hollands:2017ahy}).


\section{Isomonodromic Deformations and the Semiclassical Limit}
\label{sec:isom-tau-funct}

In the previous section, we saw that the conformal block
exponentiation \eqref{eq:sclimit6pt} effectively transforms the light
BPZ equation into a linear ODE for $\psi(z,\lambda,t)$.  The classical
conformal block $S_{\sigma}^{\pm}(\lambda,t)$ turns out to be the PVI
action.  In this section, we review the isomonodromic setup from a
Fuchsian system and how to obtain the semiclassical equations
\eqref{eq:sclimitBPZlight} and \eqref{eq:taufunctioneq} in this
approach. We will then see that \eqref{eq:taufunctioneq} is equivalent
to the definition of the isomonodromic $\tau$-function when
$(\lambda,p)$ are PVI solutions. We also make a digression about the
monodromy group and the moduli space of flat connections, summarizing
how the different objects introduced in this paper can be labeled by
the two PVI integration constants $(\sigma,s)$. Finally, we finish
this section arguing that the monodromy data of solutions of both
heavy and light BPZ equations can be encoded in a Fuchsian system. We
show that this can be consistently done only if $c=1$. This has a
two-fold purpose: first, to argue that the isomonodromic
$\tau$-function can be understood as a $c=1$ correlator and, second,
to show that the monodromy data of $c=\infty$ conformal blocks and
$c=1$ correlators can be encoded in the same Fuchsian system.


\subsection{Isomonodromic Deformations and the Garnier System}
\label{sec:isom-deform-c=1}

In this section, we recover the standard $c=1$ setup for isomonodromic
deformations of Fuchsian systems
\cite{Jimbo1981b,Jimbo:1981-2,Jimbo:1981-3,Gamayun2012}.  Let us start
with the following Fuchsian system for the vector
$\Psi(\bm{a} | z ,\lambda) = ( \psi_{1}(\bm{a}|z,\lambda), \;
\psi_{2}(\bm{a}|z ,\lambda))^{T}$
\begin{subequations}
\label{eq:fuchsian}
  \begin{align}
    \del[z]\Psi &= A(z)\Psi, \label{eq:fuchs1} \\[5pt]
    \del[\lambda]\Psi &=  -A(\lambda)\Psi,\label{eq:fuchs2}\\[5pt] 
    \del[a_{i}]\Psi &=- \left(\frac{\lambda-z}{\lambda-a_{i}} \right)
                \frac{A_{i}}{z-a_{i}}\Psi,\label{eq:fuchs3}
\end{align}
\end{subequations}
where the $A_{i}$ are $\mathfrak{gl}(2,\mathbb{C})$ matrices with
\begin{equation}
  \label{eq:38}
  \Tr A_{i} = 2\theta_{i},\quad \Tr A_{i}^{2} = 0.
\end{equation}
We call this choice of $\Tr A_{i}$ the \emph{canonical gauge}. The
Fuchsian system above has $n$ singular points
$\bm{a}=(a_{1},\ldots,a_{n})$ and corresponding monodromy coefficients
$\bm{\theta}=(\theta_{1},\ldots,\theta_{n})$ (see section
\ref{sec:moduli-space-flat} for details on the monodromy group). The
integrability conditions of \eqref{eq:fuchsian} are the
\emph{Schlesinger equations} \cite{Schlesinger1912,Gamayun2012}
\begin{subequations}
 \label{eq:schlesinger}
  \begin{align}
  \label{schl1}
    \partial_{a_{i}}A_{j}&=\frac{\lambda- a_{j}}{\lambda-a_{i}}\,
                         \frac{\left[A_{i},A_{j}\right]}{a_{i}-a_{j}},\qquad i \neq j,\qquad\\[10pt]
  \label{schl2}
    \partial_{a_{j}}A_{j}&=-\sum_{i\neq j}
                         \frac{\left[A_{i},A_{j}\right]}{a_{i}-a_{j}},\qquad
                           \partial_{\lambda}A_{j}=-\sum_{i\neq j}
                         \frac{\left[A_{i},A_{j}\right]}{\lambda-a_{i}}.
 \end{align}
\end{subequations}
These equations represent \emph{isomonodromic deformations} of the
Fuchsian system \eqref{eq:fuchsian}, as they generate a flow changing
the positions of the singular points $\bm{a}$ without changing the
monodromies. In fact, taking the trace of equations \eqref{eq:schlesinger}, it is clear that
$\Tr A_{i}$ do not change under the flow. The isomonodromic
$\tau$-function $\tau_{S}=\tau_{S}(\bm{\theta};\bm{a})$ is defined by
 \begin{equation}
   \label{eq:41}
   d\log\tau_{S} = \sum_{i < j}^{n}\Tr (A_{i}A_{j}) d\log(a_{i}-a_{j}),
 \end{equation}
 which is a closed 1-form provided that the Schlesinger equations are
 satisfied. This is the generating function of the isomonodromic
 Hamiltonians
 \begin{equation}
   \label{eq:39}
   H_{S, i} = \del[a_{i}]\log\tau_{S} = \sum_{j\neq i}\frac{\Tr( A_{i}A_{j})}{a_{i}-a_{j}},\quad i=1,\ldots,n.
 \end{equation}
 One can show from the Schlesinger equations \eqref{eq:schlesinger}
 that these Hamiltonians generate the isomonodromic flow for
 $\lambda=\lambda(\bm{a})$ \cite{Jimbo1981b,Iwasaki:1991}.

 Let us focus now on the $n=4$ case. Applying a Möbius transformation,
 we fix the singular points to $\bm{a} =(a_{0},a_{t},a_{1},a_{\infty})=(0,t,1,\infty)$ and monodromy
 parameters
 $\bm{\theta} =
 \left(\theta_{0},\theta_{t},\theta_{1},\theta_{\infty}\right)$. Then
 we have the Fuchsian system
 \begin{subequations}
\label{eq:fuchsian4}
  \begin{align}
     \del[z]\Psi &= A(z)\Psi, \label{eq:fuchs41} \\[5pt]
     \del[\lambda]\Psi &=  -A(\lambda)\Psi,\label{eq:fuchs42}\\[5pt] 
     \del[t]\Psi &=- \left(\frac{\lambda-z}{\lambda-t} \right)
                  \frac{A_{t}}{z-t}\Psi,\label{eq:fuchs43}
\end{align}
\end{subequations}
with
\begin{equation}
   \label{eq:27}
   A(z) = \sum_{i=0,1,t}\frac{A_{i}}{z-a_{i}},\quad A_{\infty} = -\sum_{i=0,1,t}A_{i}  =
   \begin{pmatrix}
     \kappa_{1}
     & 0 \\[5pt]
     0 & \kappa_{2}
   \end{pmatrix},
 \end{equation}
where $2\theta_{\infty} = \kappa_{1}-\kappa_{2}-1$ and
$ \kappa_{1}+\kappa_{2} = -2( \theta_{0}+\theta_{1}+\theta_{t} )
$. These last conditions can be solved to
\begin{equation}
  \label{eq:20}
  \kappa_{1} =\theta_{\infty}+\frac12 -\sum_{i=0,1,t}\theta_{i},\quad \kappa_{2} = -\theta_{\infty}-\frac12-\sum_{i=0,1,t}\theta_{i}.
\end{equation}
Notice that we used the SL$(2,\mathbb{C})$ gauge freedom to fix
$A_{\infty}$ to be in diagonal form. The solution $\Psi$ has
monodromies on the complex $z$-plane given by the eigenvalues of
$A_{i} \sim \text{diag}(2\theta_{i}, 0)$ (see section
\ref{sec:moduli-space-flat}). A convenient parameterization for the
$A_{i}$ was given by \cite{Jimbo:1981-2}
\begin{equation}
  \label{eq:42}
  A_{i} = \begin{pmatrix}
    p_{i}+2\theta_{i} & p_{i}q_{i} \\[5pt]
    -\frac{(p_{i}+2\theta_{i})}{q_{i}}\quad & -p_{i}
  \end{pmatrix},\quad i=0,1,t,
\end{equation}
where $p_{i}$ and $q_{i}$ are functions of $(\lambda,t)$ and the fixed
monodromy parameters $\bm{\theta}$. The diagonal form of $A_{\infty}$
in \eqref{eq:27} implies the constraints
\begin{equation}
  \label{eq:43}
  \sum_{i=0,1,t}p_{i} = \kappa_{2},\quad \sum_{i=0,1,t}p_{i}q_{i} = 0,\quad \sum_{i=0,1,t}\frac{(p_{i}+2\theta_{i})}{q_{i}} = 0.
\end{equation}
The second equation above implies that $A_{12}(z)$ must have a simple
zero in $z$ and, for consistency with 
\eqref{eq:fuchsian4}, it has to be at $z=\lambda$
\begin{equation}
  \label{eq:40}
  A_{12}(z) = k \frac{\lambda-z}{z(z-1)(z-t)},\quad k\in \mathbb{C}.
\end{equation}
We can then solve for the $q_{i}$'s in \eqref{eq:43} via
\begin{equation}
  \label{eq:44}
  p_{i}q_{i} = \res_{z=a_{i}}
  \left[
    k \frac{\lambda-z}{z(z-1)(z-t)}
  \right] = k\frac{\lambda-a_{i}}{f'(a_{i})},\quad f(z) \equiv  z(z-1)(z-t).
\end{equation}
We have only two equations left in \eqref{eq:43} for the three $p_{i}$'s, so we introduce
the variable
\begin{equation}
  \label{eq:45}
 \mu= \sum_{i=0,1,t}\frac{p_{i}+2\theta_{i}}{\lambda-a_{i}}
\end{equation}
and solve the constraints for them in terms of $(\lambda,\mu,t)$.
The expressions for the $p_{i}$'s are not particularly enlightening to
display here and can be found in \cite{Jimbo:1981-2}. 

Let us consider the second order ODE for the first component of
$\Psi = (\psi_{1}\,,\, \psi_{2})^{T}$
\begin{equation}
  \label{eq:46}
  \del[z]^{2}\psi_{1} -(\Tr A(z) + \del[z]\log A_{12}(z))\del[z]\psi_{1} + 
  \left(
    \det A(z) + A_{11}(z)\del[z]\log\frac{A_{12}(z)}{A_{11}(z)}
  \right)\psi_{1} = 0.
\end{equation}
Writing $(p_{i},q_{i})$ in terms of $(\lambda,\mu,t)$ and using the
parameterization \eqref{eq:42} above, we get the deformed Heun
equation in canonical form
\begin{subequations}
\label{eq:garnier}  
\begin{align}
&\del[z]^{2}\psi_1+g_{1}(z)\del[z]\psi_1+g_{2}(z)\psi_1=0, \\[10pt]
g_{1}(z)&=\frac{1-2\theta_0}{z}+\frac{1-2\theta_1}{z-1}+
\frac{1-2\theta_t}{z-t}-\frac{1}{z-\lambda}, \\[10pt]
g_{2}(z)&=\frac{\kappa_1(\kappa_2+1)
}{z(z-1)}-\frac{t(t-1)K}{z(z-1)(z-t)}+
\frac{\lambda(\lambda-1)\mu}{z(z-1)(z-\lambda)},
\end{align}
\end{subequations}
with the accessory parameter $K=  K(\bm{\theta}; \lambda,\mu,t ) $ given by
\begin{multline}
  \label{eq:kamiltonian}
  K(\bm{\theta};\lambda,\mu,t) 
=
  \frac{\lambda(\lambda-1)(\lambda-t)}{t(t-1)}
  \left[\mu^2-\left(\frac{2\theta_0}{\lambda}+\frac{2\theta_1}{\lambda-1}+
      \frac{2\theta_t-1}{\lambda-t}\right)\mu+\frac{\kappa_{1}(\kappa_{2}+1)}{\lambda(\lambda-1)}\right].
\end{multline}
Notice that \eqref{eq:20} implies in
\begin{equation}
  \label{eq:68}
  \kappa_{1}(\kappa_{2}+1) = 
  \left(
    \sum_{i=0,1,t}\theta_{i}-\frac12 
  \right)^{2}-\theta_{\infty}^{2}.
\end{equation}
As we saw above, the integrability conditions for the Fuchsian system
are equivalent to the isomonodromic equations
\eqref{eq:schlesinger}. Using the parameterization above in terms of
$(\lambda,\mu,t)$, the isomonodromic equations reduce to the
\emph{Garnier system}~\cite{Garnier:1912,Iwasaki:1991}
\begin{equation}
  \label{eq:18}
  \dot{\lambda} = \Dpfrac[]{K}{\mu},\quad   \dot{\mu} = -\Dpfrac[]{K}{\lambda}.
\end{equation}
We denote a solution of the isomonodromic flow as
$(\lambda(t),\mu(t))$.  The second order equation for $\lambda(t)$ is
the Painlevé VI equation \cite{Jimbo:1981-2}. The PVI solutions are,
in general, transcendental, i.e., cannot be reduced to simple
algebraic or special functions.  Jimbo has used the isomonodromic
technique \cite{Jimbo:1982} to find the asymptotics of the
$\tau$-function \eqref{eq:39} and, consequently, of the PVI
transcendents, near its critical points $t=0,1,\infty$. The full
expansion of the PVI $\tau$-function was found only recently in
\cite{Gamayun2012}. We will review this formula in section
\ref{sec:access-param-from}.

\subsection{Semiclassical BPZ Equations from the Fuchsian System}
\label{sec:semicl-bpz-from}

We claimed in section \ref{sec:isom-semicl-limit} that a certain heavy-light
6-point correlator naturally encodes isomonodromic equations. To show
this explicitly, we  obtain the semiclassical Fuchsian
equation \eqref{eq:sclimitBPZlight} from the Fuchsian system
\eqref{eq:fuchsian4}. 
Applying the transformation
\begin{equation}
  \label{eq:50}
  \psi_{1}(t|z,\lambda) =(z-\lambda)^{\frac{1}{2}}
  \prod_{i=0,1,t}(z-a_{i})^{-\frac{1}{2}+\theta_{i}}\psi(z,\lambda,t) 
\end{equation}
to \eqref{eq:garnier}, we find the semiclassical Fuchsian equation
\begin{multline}
  \label{eq:47}
  \del[z]^{2}\psi + \left( -\frac{t(t-1)H}{z(z-1)(z-t)}+
    \frac{\lambda(\lambda-1)p}{z(z-1)(z-\lambda)} +\right.\\[5pt]
  \left. +\sum_{i=0,1,t} \frac{\delta_{i}}{(z-a_{i})^{2}} +
    \frac{-\tfrac34}{(z-\lambda)^{2}} +
    \frac{\delta_{\infty}-\sum_{i=0,1,t}\delta_{i}+\tfrac34}{z(z-1)}
  \right)\psi = 0,
\end{multline}
where the monodromy parameters are encoded by
\begin{equation}
  \label{eq:deltas}
  \delta_{i} \equiv \delta(\theta_{i}) = \frac{1}{4} - \theta_{i}^{2}
\end{equation}
and the accessory parameters are
\begin{align}
  \label{eq:56}
  p &= \mu + \sum_{i=0,1,t}\frac{1-2\theta_{i}}{2(\lambda-a_{i})},\\[10pt]
  \label{eq:48}
  H(\bm{\theta};\lambda,p,t ) &=\frac{\lambda(\lambda-1)(\lambda-t)}{t(t-1)} \left[ p^{2}- \left(
                     \frac{1}{\lambda}+\frac{1}{\lambda-1} \right)p +\right.\nonumber
  \\[5pt]
    &\left. \qquad +
      \frac{\delta_{0}}{\lambda^{2}}+\pfraction{2}{\lambda}{1}{\delta_{1}}+\pfraction{2}{\lambda}{t}{\delta_{t}}+\frac{\delta_{\infty}-\delta_{0}-\delta_{1}-\delta_{t}+\tfrac34}{\lambda(\lambda-1)}      \right].
\end{align}
Now comparing \eqref{eq:47} and \eqref{eq:48}
with \eqref{eq:sclimitBPZlight} and \eqref{eq:sclimitBPZheavy}, we
find
\begin{equation}
  \label{eq:55}
  H = -\del[t]S_{\sigma}^{\pm}(\lambda,t) ,\quad p = \del[\lambda]S_{\sigma}^{\pm}(\lambda,t),
\end{equation}
and thus we have recovered the semiclassical BPZ equations of section
\ref{sec:isom-semicl-limit} from the isomonodromic Fuchsian system.

At this point, probably it is not clear to the reader what is the
relation between the classical intermediate momentum $\sigma$ in the
semiclassical action and the Fuchsian system parameters
$(\lambda,\mu,t)$ (or $(\lambda,p,t)$ in the Fuchsian equation
\eqref{eq:47}). We will clarify this point in the rest of this section
by discussing the relationship between the monodromy group, the moduli
space of flat connections and the semiclassical action.

\subsection{Monodromy Group, Flat Connections
  and the Semiclassical Action}
\label{sec:moduli-space-flat}

\begin{figure}
  \centering
  \includegraphics[width=.5\textwidth]{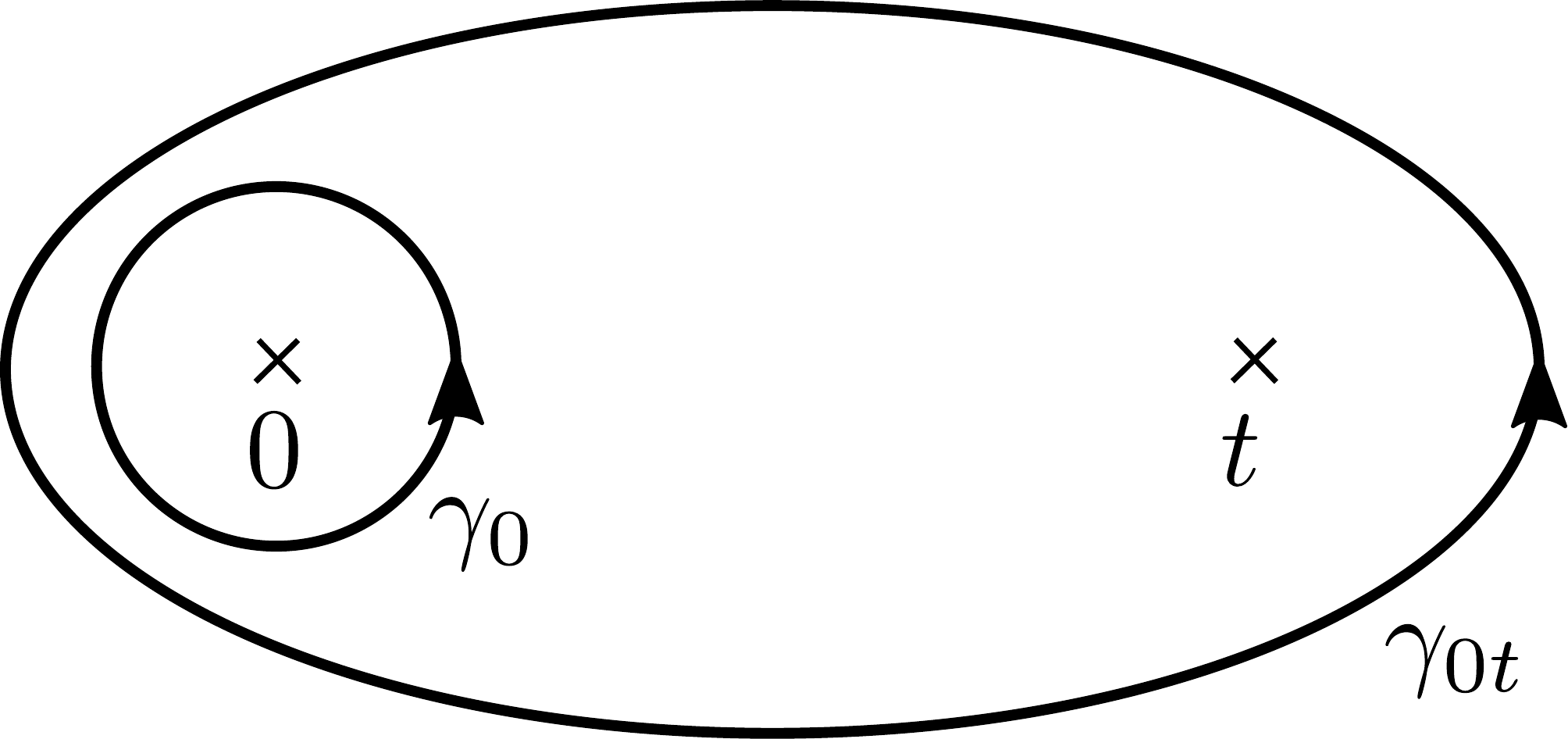}
  \caption{An elementary path $\gamma_0$ has monodromy $\theta_0$ and a
    composite path $\gamma_{0t}$ enclosing two singular points has
    composite monodromy $\sigma_{0t}$.}
  \label{fig:composite_mono}
\end{figure}

In this section, we will consider the Fuchsian system
\eqref{eq:fuchsian4} in the $\slc$ gauge, where $\Tr A_{i} = 0$. This
can be obtained by the gauge transformation \eqref{eq:86} discussed
below. Assuming that $\lambda$ and $t$ are fixed, the formal solution
of \eqref{eq:fuchsian4} is given by
\begin{equation}
  \label{eq:19}
  \Psi(z) = \mathcal{P}\exp
  \left(
    \int^{z}A 
  \right)\Psi(z_{0}),
\end{equation}
where $\mathcal{P}$ represents a path-ordered exponential and $z_{0}$
is an arbitrary base point.  A consequence of this formula is that the
poles of the gauge connection $A(z)$ correspond to branch points of
$\Psi(z)$. If we do the analytic continuation of $\Psi(z)$ around a
closed path $\gamma$, enclosing one or more singular points, the
solution will change by a \emph{monodromy matrix} $M_{\gamma}$, i.e.,
$\Psi_{\gamma} = M_{\gamma}\Psi$.  Elementary paths enclosing only one
singular point $a_{i}$ have monodromy matrix $M_{i}$ and we label
those matrices by their trace $\Tr M_{i} = 2\cos(2\pi\theta_{i})$. The
four-point monodromy group is then generated by three out of four
$\slc$ matrices obeying the monodromy identity
\begin{equation}
  \label{eq:4}
  M_{0}M_{t}M_{1}M_{\infty} = \mathbbold{1}.
\end{equation} 
As $\Psi(z)$ is analytic everywhere except at the branch cuts, the
knowledge of the related monodromy data essentially determines the
solution next to these points. The global information on how to
connect different local solutions is encoded in the composite
monodromies, obtained when a path encloses two singular points (two or
more points for $n > 4$). For example,
$\Psi_{\gamma_{0t}} = M_{0}M_{t}\Psi$ has composite monodromy
parameter defined by $\Tr M_{0}M_{t} = -2\cos(2\pi\sigma_{0t})$ (see
figure \ref{fig:composite_mono}).
All representations of the 4-point monodromy group are labeled by 4
elementary monodromies
$\bm{\theta}=(\theta_{0},\theta_{t},\theta_{1},\theta_{\infty})$ and 3
composite monodromies
$\bm{\sigma}=(\sigma_{0t},\sigma_{1t},\sigma_{01})$, as this is the
sufficient data to generate all possible loops around singular points
in the 4-punctured sphere. We call $(\btheta;\bm{\sigma})$ the
\emph{monodromy data} associated to $\Psi(z)$.

Let us define the monodromy parameters by
\begin{equation}
  \label{eq:monodromyparameters}
  p_i=\Tr M_i = 2\cos2\pi\theta_{i}, \quad p_{ij}=\Tr M_i M_j = 2\cos2\pi\sigma_{ij},\quad i,j =0,1,t,\infty.
\end{equation}
Assuming that the $p_{i}$'s are fixed, irreducible representations of
the monodromy group are labeled by three composite monodromies
$(p_{0t},p_{1t},p_{01})$.  For $\slc$ matrices, the
monodromy parameters also obey the Fricke-Jimbo relation
\cite{Jimbo:1982}
\begin{multline}
  p_{0t}p_{1t}p_{01}+p_{0t}^2+p_{1t}^2+p_{01}^2+p_0^2+p_t^2+p_1^2+
  p_0p_1p_tp_\infty = \\
  (p_0p_t+p_1p_\infty)p_{0t}+(p_1p_t+p_0p_\infty)p_{1t}
  +(p_0p_1+p_tp_\infty)p_{01}+4,
\label{eq:frickejimbo}
\end{multline} 
and thus only two composite monodromies are independent of each other.
Following \cite{Jimbo:1982,Iorgov2015}, if we fix the $\btheta$ and
$\sigma_{0t}$, the Fricke-Jimbo relation can be parametrized in terms
of $s_{0t}$ as
\begin{subequations}
\label{Ji_par1}
\begin{align}
&\left(p_{0t}^2-4\right)p_{1t}=\,D_{t,+}\, s_{0t}\,+D_{t,-}\, s_{0t}^{-1}\,+D_{t,0},\\
&\left(p_{0t}^2-4\right)p_{01}=D_{u,+}\, s_{0t}+D_{u,-}\, s_{0t}^{-1}+D_{u,0},
\end{align}
\end{subequations}
with coefficients given by
\begin{subequations}
\label{Ji_par2}
\begin{align}
  &D_{t,0}=p_{0t}\left(p_0p_1+p_tp_\infty\right)-2\left(p_0p_\infty+p_tp_1\right),\\[5pt]
  &D_{u,0}=p_{0t}\left(p_{t}p_1+p_0p_\infty\right)-2\left(p_0p_1+p_tp_\infty\right),\\[5pt]
  &D_{t,\pm}=16\prod_{\epsilon=\pm}\sin\pi\left(\theta_{t}\mp \sigma_{0t}+\epsilon \theta_{0}\right)\sin\pi\left(\theta_{1}\mp \sigma_{0t}+\epsilon \theta_{\infty}\right),\\[5pt]
  &D_{u,\pm}=-D_{t,\pm}e^{\mp 2\pi i \sigma_{0t}}.
\end{align}
\end{subequations}
Solving the system \eqref{Ji_par1} for $s_{0t}$ when
\begin{equation}
  \label{eq:sigmacond}
  \sigma_{ij} +\epsilon \theta_{i}+\epsilon' \theta_{j} \notin
  \mathbb{Z} ,\quad \epsilon,\epsilon' = \pm 1,
\end{equation}
we get
\begin{multline}
\label{eq:sparameter}
s_{0t}^{\pm}(\cos 2\pi(\theta_t\mp\sigma_{0t})-\cos 2\pi\theta_0) (\cos
2\pi(\theta_1\mp \sigma_{0t})-\cos \pi\theta_\infty) \\
=(\cos 2\pi\theta_t\cos 2\pi\theta_1+\cos 2\pi\theta_0 \cos
2\pi\theta_\infty\pm i\sin 2\pi\sigma_{0t}\cos 2\pi\sigma_{01}) \\
-(\cos 2\pi\theta_0\cos 2\pi\theta_1+\cos 2\pi\theta_t\cos
2\pi\theta_\infty\mp i\sin 2\pi\sigma_{0t}\cos 2\pi\sigma_{1t})e^{\pm
  2 \pi i\sigma_{0t}}.
\end{multline}  
The special cases when \eqref{eq:sigmacond} is not true correspond to
reducible representations, which are all listed in the context of PVI
solutions in \cite{Guzzetti2012b}. In those cases, to find $s$ we
should go back to the Fricke-Jimbo relation \eqref{eq:frickejimbo}.
In conclusion, we can label irreducible representations of the 4-point
monodromy group by two parameters $(\sigma_{0t},s_{0t})$.  This
parametrization is essentially the same under the permutation of the
composite monodromies and its related to the number of independent
ways to slice the 4-punctured sphere into two pairs of pants
\cite{Iorgov2015}.

\subsubsection{Summary of Parameters}
\label{sec:summary-parameters}

In section \ref{sec:isom-deform-c=1}, we parametrized the gauge
connection $A(z)$ in terms of elementary monodromies $\btheta$ and two
extra parameters $(\lambda,\mu)$ (or $(\lambda,p)$). Therefore, for
fixed $t$, the \emph{moduli space }of flat connections $A(z)$ can be
labeled by $(\btheta; \lambda,\mu)$. From \eqref{eq:19}, it is clear
that there should be a map between the moduli parameters
$(\lambda,\mu)$ and the monodromy parameters
$(\sigma_{0t},s_{0t})$. This is the Riemann-Hilbert map for Fuchsian
systems, the map between irreducible representations of the monodromy
group and the moduli space of flat connections
\cite{Iwasaki:1991,Iorgov2015}. As explicitly shown in
\cite{Jimbo:1982}, isomonodromic deformations define such map via the
integration constants $(\sigma_{0t},s_{0t})$. Jimbo obtained the
asymptotics of $A(z;\lambda(t),\mu(t),t)$ for small $t$ and showed
that the formulas only depend on $(\btheta;\sigma_{0t},s_{0t})$. This
can then be used to find the asymptotics of the $\tau$-function and
the PVI solutions $(\lambda(t),\mu(t))$ in terms of
$(\btheta;\sigma_{0t},s_{0t})$. This will become clear in section
\ref{sec:access-param-from} when we discuss the isomonodromic
$\tau$-function expansion. 

The semiclassical equation \eqref{eq:47} is parametrized by $\lambda$
and the accessory parameters $H$ and $p$, for fixed $t$ and
$\btheta$. As $H$ is a function of $(\lambda,p,t)$, these parameters
label the possible equations. If we assume that $(\lambda(t),p(t))$
are solutions of the isomonodromic equations, i.e.
\begin{equation}
  \label{eq:garniersystem}
  \dot{\lambda} = \Dpfrac{H}{p},\quad \dot{p} = -\Dpfrac{H}{\lambda},
\end{equation}
then $(\lambda(t),p(t))$ are PVI solutions and can be labeled by the
monodromy parameters $(\sigma_{0t},s_{0t})$. Therefore, for fixed $(t,\btheta)$, we have
that $(\lambda(t),p(t)) \sim (\sigma_{0t},s_{0t})$.

We also showed in \eqref{eq:55} that the accessory parameters are
given by derivatives of the semiclassical action
$S_{\sigma}(\lambda(t),t)$.  To complete the picture relating the
semiclassical equation \eqref{eq:47} and the monodromy parameters, we need to show that
the classical intermediate momentum $\sigma$ is the \emph{composite
  monodromy} $\sigma_{0t}$. Let us assume that, for 
fixed $\lambda \neq 0,1,t,\infty$, the solution $\psi(z,\lambda,t)$ of
\eqref{eq:47} has a small $t$ expansion as
\begin{equation}
  \label{eq:21}
  \psi(z,\lambda,t) = z^{\frac{1}{2}-\sigma_{0t}} f(z,\lambda) + \mathcal{O}(t).
\end{equation}
Assuming that \eqref{eq:21} is well-defined at both $z=0$ and $z=t$,
for $t$ small enough, $\sigma_{0t}$ represents the composite monodromy
parameter.  Substituting \eqref{eq:21} in \eqref{eq:47}, we find at
lowest order
\begin{equation}
  \label{eq:22}
  H_{0} = \delta_{0}+\delta_{t}-\delta_{\sigma_{0t} }
\end{equation}
where $H_{0} = \lim_{t\rightarrow 0}t(1-t) H$. From CFT, we know that
for small $t$
\begin{equation}
  \label{eq:24}
  S_{\sigma}(\lambda,t) \sim (\delta_{\sigma
  }-\delta_{0}-\delta_{t})\log(t) 
\end{equation}
and thus
\begin{equation}
  \label{eq:61}
  H(\lambda,p,t) = -\del[t]S_{\sigma}(\lambda,t)  \sim\frac{  \delta_{0}+\delta_{t}-\delta_{\sigma }}{t},
\end{equation}
which agrees with \eqref{eq:22} if $\sigma=\sigma_{0t}$. From here and
in the rest of the paper, we define
$(\sigma,s)\equiv(\sigma_{0t},s_{0t})$, unless otherwise stated.

In summary, for fixed $(t,\btheta)$, we have the following set of parameters
\begin{subequations}
  \begin{align*}
    \text{Monodromy representations:} &\quad (\sigma,s)\\
   \text{Semiclassical Action:}&\quad (\lambda,\sigma)\\
 \text{Flat Connection:}&\quad (\lambda,\mu)\quad (\text{or}\,
                             (\lambda,p))
  \end{align*}
\end{subequations}
The assumption that $(\lambda(t),\mu(t))$ is a solution of the
isomonodromic flow connects the different parameters, since all
quantities of interest can be phrased in terms of $(\sigma,s)$.

\subsection{Fuchsian System and $c=1$ BPZ equations}
\label{sec:fuchsian-system-c=1}

A natural question is whether it is possible to encode the level-2
heavy and light BPZ equations into a Fuchsian system for any value of
$c$. This was proved for a single $c=1$ BPZ equation
in~\cite{Novikov2009}. As we show below, we can recover each level-2
BPZ equation separately for arbitrary $c$ from an appropriate Fuchsian
system. However, we can only consistently recover both BPZ equations
if $c=1$.  The relationship is true if the associated linear system
allows for isomonodromic deformations.

Let us first change the gauge of \eqref{eq:fuchsian} by applying the
transformation
\begin{equation}
  \label{eq:86}
  \Psi = \prod_{i=1}^{n}[(z-a_{i})(\lambda-a_{i})]^{\theta_{i}}\Phi,\quad A_{i} = B_{i}+\theta_{i}\mathbb{1}
\end{equation}
to \eqref{eq:13}, we get 
\begin{subequations}
\label{eq:13}
\begin{align}
  \epsilon_{1}\del[z]\Phi &= B(z)\Phi, \label{eq:30} \\[5pt]
 \epsilon_{2}\del[\lambda]\Phi &=  B(\lambda)\Phi,\label{eq:35}\\[5pt] 
\alpha\del[a_{i}]\Phi &= \left(\frac{\lambda-z}{\lambda-a_{i}} \right)
                \frac{B_{i}}{z-a_{i}}\Phi,\quad (i=1,\ldots,n)\label{eq:34}
\end{align}
\end{subequations}
where $\epsilon_{1},\epsilon_{2}$ and $\alpha$ are three arbitrary
constants and
\begin{gather}
  \label{eq:15}
  B(z) = \sum_{i=1}^{n}\frac{B_{i}}{z-a_{i}} =
  \begin{pmatrix}
    B_{11}(z) & B_{12}(z)\\
    B_{21}(z) & B_{22}(z)
  \end{pmatrix}
,\quad \sum_{i=1}^{n}B_{i} = 0,
\end{gather}
such that
\begin{gather}
  \label{eq:16}
  \Tr B_{i} = 0,\quad \Tr B_{i}^{2} = -2\epsilon_{1}\epsilon_{2}\Delta_{i}.
\end{gather}
The choice of $\Tr B_{i}$ sets the $\slc$ gauge for the Fuchsian
system.

To derive a second order equation for $\Phi$, we take the derivative of
\eqref{eq:30} with respect to $z$
\begin{equation}
  \label{eq:31}
  \epsilon_{1}\del[z]^{2}\Phi = (\del[z]B + \frac{B^{2}}{\epsilon_{1}})\Phi.
\end{equation}
Using the relation for $\mathfrak{sl}(2,\mathbb{C})$ matrices
\begin{equation}
  \label{eq:auxeqs}
  B_{i}B_{j}+B_{j}B_{i} = \Tr(B_{i}B_{j})\,\mathbb{1},
\end{equation}
we can easily show that
\begin{equation}
  \label{eq:32}
  B^{2} = \sum_{i=1}^{n} 
  \left(
    \frac{-\epsilon_{1}\epsilon_{2}\Delta_{i}}{(z-a_{i})^{2}}+\frac{H_{i}}{z-a_{i}}
  \right)\mathbb{1},
\end{equation}
where the accessory parameters are defined by
\begin{equation}
  \label{eq:33}
  H_{i}\equiv \sum_{j\neq i} \frac{\Tr (B_{i}B_{j})}{a_{i}- a_{j}}.
\end{equation}
From \eqref{eq:34} and \eqref{eq:35}, we obtain
\begin{equation}
  \label{eq:36}
  \sum_{i=1}^{n}\frac{B_{i}}{(z-a_{i})^{2}} = \frac{-\epsilon_{1}\del[z]\Phi+\epsilon_{2}\del[\lambda]\Phi}{z-\lambda}+\sum_{i=1
}^{n}\frac{\alpha\del[a_i]\Phi}{z-a_{i}}.
\end{equation}
Plugging \eqref{eq:32} and \eqref{eq:36} into \eqref{eq:31} and
dividing by $\epsilon_{2}$, we get
\begin{equation}
  \label{eq:37}
  \frac{\epsilon_{1}}{\epsilon_{2}}
  \left(
    \del[z]^{2}\Phi - \frac{1}{z-\lambda}\del[z]\Phi
  \right) +\frac{\del[\lambda]\Phi}{z-\lambda}+\sum_{i=1}^{n}
  \left(
    \frac{\Delta_{i}}{(z-a_{i})^{2}}+\frac{\frac{\alpha}{\epsilon_{2}}\del[a_i]-\frac{1}{\epsilon_{1}\epsilon_{2}}H_{i}}{z-a_{i}}
  \right)\Phi =0.
\end{equation}
Now we apply the transformation
\begin{equation}
  \label{eq:49}
  \Phi(\bm{a}| z,\lambda) =
  (z-\lambda)^{\frac{1}{2}}\,[h(\bm{a})]^{\frac{1}{\alpha\epsilon_{1}}}\,\chi(\bm{a}| z,\lambda)
\end{equation}
in (\ref{eq:37}), with the choice
\begin{equation}
  \label{eq:83}
   H_{i}(\bm{a}) = \del[a_i]\log h(\bm{a}).
\end{equation}
Notice that \eqref{eq:33} and \eqref{eq:39} imply that $h(\bm{a})$ is
the $\tau$-function up to a overall function of $t$. Then we get the
following equation for $\chi$
\begin{equation}
  \label{eq:BPZfromLinear}
  \left[
    \frac{\epsilon_{1}}{\epsilon_{2}}\del^{2} +
    \pfraction{2}{z}{\lambda}{\Delta_{H}}+
    \pfraction{}{z}{\lambda}{\del[\lambda]}+\sum_{i=1}^{n}
    \left(
      \pfraction{2}{z}{a_{i}}{\Delta_{i}}+\pfraction{}{z}{a_{i}}{\frac{\alpha}{\epsilon_{2}}\del[a_{i}]}
    \right)
  \right]\chi(\bm{a}| z,\lambda)=0,
\end{equation}
where $\Delta_{H} =
-\frac{3\epsilon_{1}}{4\epsilon_{2}}-\frac{1}{2}$. Setting
$\alpha = \epsilon_{2}$ and $b^{2} = \epsilon_{2}/\epsilon_{1}$, this
equation becomes the generalization of the BPZ equation
\eqref{eq:BPZa} for a correlator with $n$ arbitrary insertions at
$\bm{z}=\bm{a}$ and two degenerate insertions, one at $z$ and another
at $\lambda$. However, if we repeat the same procedure for
\eqref{eq:35}, we only get the second BPZ equation \eqref{eq:BPZb} if
we choose $\alpha=-\epsilon_{1}$. This means that
$\alpha=\epsilon_{2} = -\epsilon_{1}$ is a sufficient condition for the
Fuchsian system (\ref{eq:13}) consistently reproduce the two BPZ
equations.  Accordingly, $b=i$ and thus $c=1$ in the CFT
interpretation. Using the parametrization \eqref{eq:1} and
\eqref{eq:2}, the $c=1$ BPZ equations are thus
\begin{subequations}
\label{eq:BPZ}
  \begin{align}
    \left[
    -\del^{2} +
    \pfraction{2}{z}{\lambda}{\frac{1}{4}}+
    \pfraction{}{z}{\lambda}{\del[\lambda]}+\sum_{i=1}^{n}
    \left(
    \pfraction{2}{z}{a_{i}}{\theta_{i}^{2}}+\pfraction{}{z}{a_{i}}{\del[a_{i}]}
    \right)
    \right]\chi(\bm{a}|z,\lambda)&=0, \label{eq:BPZc1a}\\[10pt]
    \left[-\del[\lambda]^{2} +
    \pfraction{2}{\lambda}{z}{\frac{1}{4}}+
    \pfraction{}{\lambda}{z}{\del[z]}+\sum_{i=1}^{n}
    \left(
    \pfraction{2}{\lambda}{a_{i}}{\theta_{i}^{2}}+\pfraction{}{\lambda}{a_{i}}{\del[a_{i}]}
    \right)
    \right]\chi(\bm{a}|z,\lambda) \label{eq:BPZc1b}&=0,
\end{align}
\end{subequations}
where $\Delta_{i} = \theta_{i}^{2}$ is the $c=1$ conformal weight.
This shows that the Fuchsian system \eqref{eq:13} with
$\epsilon_{2}=-\epsilon_{1}$ simultaneously encodes the monodromy data
of the two $c=1$ BPZ equations above. Moreover, this fact is the
starting point of the original argument of why the isomonodromic
$\tau$-function is equivalent to a $c=1$ correlator
\cite{Gamayun2012}. For a proof of this fact, see \cite{Iorgov2015}.

\section{Accessory parameter from the Isomonodromic $\tau$-function}
\label{sec:access-param-from}
We previously discussed how the accessory parameter
$H(\lambda(t),p(t),t)$ in \eqref{eq:47} can be written in terms of the
monodromy data $(\bm\theta;\sigma,s)$, given that $(\lambda(t),p(t))$
are solutions of the isomonodromic flow. Now we want to use this fact
to write the accessory parameter expansion $H_{x}$ of Heun's equation
\eqref{eq:heun} in terms of the monodromy data. The key point is to
impose a special initial condition for the isomonodromic flow, which
we discuss next. Then we will present our algorithm on how to solve
this initial condition and obtain $H_{x}$ using the $\tau$-function
expansion of \cite{Gamayun2012}. We review the $\tau$-function
expansion in section \ref{sec:cft-expansion-tau}, solve the initial
condition in section \ref{sec:solving-lambdax=x} and finally present
the accessory parameter expansion in section
\ref{sec:access-param-expans}.

\subsection{Accessory Parameter from Initial Conditions}
\label{sec:isom-equat-c=1}


Let us focus on the $n=4$ isomonodromic system \eqref{eq:fuchsian4}.
Here we discuss how to recover the accessory parameter of Heun's
equation \eqref{eq:heun} from the deformed Heun's equation
\eqref{eq:garnier} by an appropriate initial condition on
$(\lambda(t),\mu(t))$ at $t=x$.
The canonical form of \eqref{eq:heun}, obtained by the transformation
$\psi = \prod_{i=0,1,x}(z-a_{i})^{(1-2\theta_{i})/2}\,y$, is given by
\begin{equation}
  \label{eq:heuncanonical}
  y^{\prime\prime}(z) +
  \left(\frac{1-2\theta_0}{z} + \frac{1-2\theta_1}{z-1} +
    \frac{1-2\theta_x}{z-x} \right) y^{\prime}(z) +
  \left(\frac{\kappa_1(\kappa_2+1)}{z(z-1)}-\frac{x(x-1)K_x}{z(z-1)(z-x)}
  \right) y(z) = 0,
\end{equation} 
where the accessory parameters of both equations are related by
\begin{equation}
  \label{eq:htok}
    H_{x} = K_{x} +\frac{(1-2\theta_{0})(1-2\theta_{x})}{2x}+\frac{(1-2\theta_{1})(1-2\theta_{x})}{2(x-1)}.
\end{equation}
Consider the deformed Heun equation \eqref{eq:garnier}, which we
repeat here for convenience to the reader,
\begin{multline}
  \label{eq:23}
  y''+ \left( \frac{1-2\theta_0}{z}+\frac{1-2\theta_1}{z-1}+
    \frac{1-2\theta_t}{z-t}-\frac{1}{z-\lambda(t)}
  \right)y'+\\[10pt]
  +\left( \frac{\kappa_1(\kappa_2+1)
    }{z(z-1)}-\frac{t(t-1)K(\btheta;\lambda(t),\mu(t),t)}{z(z-1)(z-t)}+
    \frac{\lambda(\lambda-1)\mu(t)}{z(z-1)(z-\lambda(t))} \right)y=0,
\end{multline}
with
\begin{multline}
  \label{eq:kamiltonian}
  K(\bm{\theta};\lambda,\mu,t) 
=
  \frac{\lambda(\lambda-1)(\lambda-t)}{t(t-1)}
  \left[\mu^2-\left(\frac{2\theta_0}{\lambda}+\frac{2\theta_1}{\lambda-1}+
      \frac{2\theta_t-1}{\lambda-t}\right)\mu+\frac{\kappa_{1}(\kappa_{2}+1)}{\lambda(\lambda-1)}\right].
\end{multline}
We recover \eqref{eq:heuncanonical} from \eqref{eq:23} by applying the
following initial condition for the isomonodromic flow to \eqref{eq:23}
\cite{Novaes2014c,daCunha:2015fna} 
\begin{equation}
  \label{eq:initialcondition}
  t=x,\quad\lambda(x) = x, \quad \mu(x) = -\frac{K_{x}}{2\theta_{t}},\quad \theta_{t} = \theta_{x}-\frac12.
\end{equation}
We show in Appendix \ref{sec:well-posedn-lambd} that this condition is well-posed with respect to the isomonodromic
equations. As we
discussed before, $(\lambda(t),\mu(t))$ have $(\sigma,s)$ as
integration constants, so imposing \eqref{eq:initialcondition} seems
equivalent to fix the values of $(\sigma,s)$ separately. However, as
we are going to see next, due to the special structure of the
isomonodromic flow, the condition $\lambda(x)=x$ implies a non-trivial
relation $s=s(\sigma,x)$ between the composite monodromy
parameters. This means that Heun's equation \eqref{eq:heuncanonical}
lives in a proper subspace of all possible Fuchsian opers with 4
singulars points.

Litvinov \emph{et al} \cite{Litvinov:2013sxa} found the accessory
parameter expansion of $H_{x}$ by regularizing the PVI action via the
solution of PVI equation with $\lambda(x)=\infty$. Here we present an
alternative way to calculate the accessory parameter using the
isomonodromic $\tau$-function expansion \cite{Gamayun2012} with the
condition $\lambda(x)=x$. This gives a clearer implementation of the
proposal \cite{Litvinov:2013sxa} and, in addition, suggests a deeper
connection between $c=\infty$ and $c=1$ conformal blocks.  Notice that
the condition $\lambda(x)=\infty$ is equivalent to our choice by a
bi-rational transformation \cite{Guzzetti2012b}. 

Our algorithm to find the accessory parameter expansion consists of
three steps
\begin{enumerate}
\item Write the accessory parameter $H_{x}$ in terms of the
  isomonodromic $\tau$-function;
\item Solve the initial condition $\lambda(x)=x$ to obtain a monodromy
  constraint $s=s(\sigma,x)$;
\item Substitute the constraint $s=s(\sigma,x)$ in the $\tau$-function
  to obtain $H_{x}$.
\end{enumerate}
We will show below that the condition $\lambda(x)=x$ can be phrased in
terms of the $\tau$-function, which is essential to find $s(\sigma,x)$
as a series expansion in $x$. We remind the reader that the
well-posedness of the initial condition is discussed in appendix
\ref{sec:well-posedn-lambd}. After obtaining $s(\sigma,x)$, to obtain
the accessory parameter expansion is just a matter of straightforward
calculation. This calculation is cumbersome at higher orders, so we
show it explicitly only up to order $x^{0}$ in this section.  In
Appendix \ref{sec:numer-checks-access}, we reproduce the analytic CFT
formulas up to order $x^{2}$ and present numerical evidence up to
order $x^{5}$, compared to the direct CFT approach.

The key to understand the relation $s=s(\sigma,x)$ is to phrase the
initial conditions in terms of the isomonodromic $\tau$-function. We
will do this in the next subsection. Here we express $H_{x}$ in terms
of the $\tau$-function.  For convenience we define
\begin{equation}
  \label{eq:69}
  \bm{\theta}_{s_{1},s_{2}} = (\lmtheta{\tfrac{1}{2} s_{1}}{\tfrac{1}{2} s_{2}}),\quad s_{1},s_{2}=0,\pm,
\end{equation}
and $\bm{\theta}_{0,0}\equiv \bm{\theta}$.  The definition of the
isomonodromic Hamiltonian \eqref{eq:39} for our particular 4-point
case is given by
\begin{equation}
  \label{eq:65}
  H_{S,t} = \frac{\Tr(A_{0}A_{t})}{t}+\frac{\Tr(A_{1}A_{t})}{t-1}.
\end{equation}
Using the parameterization \eqref{eq:42} for the $A_{i}$'s in
\eqref{eq:65}, we find that
\begin{equation}
  \label{eq:isocanonical}
  H_{S,t}(\bm{\theta}_{0,+};\lambda,\mu,t) = K(\bm{\theta}_{+,+}
  ;\lambda,\mu,t)+ 
\frac{4\theta_{0}\theta_{t}}{t}  + \frac{4\theta_{1}\theta_{t}}{t-1}. 
\end{equation}
Imposing the initial conditions \eqref{eq:initialcondition} in
\eqref{eq:isocanonical}, we obtain
  \begin{align}
    K_{x} &= \left.\frac{d}{dt}\log[t^{-4\theta_{0}\theta_{t}} (1-t)^{-4\theta_{1}\theta_{t}}
            \tau_{S}(\bm{\theta}_{0,+}
            ;t )]\right|_{t=x}
            \label{eq:litvinov_cond2},
  \end{align}
where we used the $\tau$-function definition 
\begin{equation}
  \label{eq:8}
  H_{S,t}(\bm{\theta}_{0,+};\lambda(t),\mu(t),t) =  \Dfrac{}{t}\log 
   \tau_{S}(\bm{\theta}_{0,+}
  ;t).
%
\end{equation}
Notice that \eqref{eq:litvinov_cond2} substitutes the initial
condition for $\mu(x)$ in \eqref{eq:initialcondition}, as it gives
$K_{x}$ in terms of the monodromy data. Together with \eqref{eq:htok},
we can then find $H_{x}$ in terms of the $\tau$-function.

\subsection{CFT Expansion of the $\tau$-function}
\label{sec:cft-expansion-tau}

In order to proceed, we need to introduce the $\tau$-function
expansion of \cite{Gamayun2012}.  For later convenience, we define a
slightly different $\tau$-function
\begin{equation}
  \label{eq:66} \tau_{S}(\bm{\theta};t)= t^{2\theta_{0}\theta_{t}}(1-t)^{4\theta_{1}\theta_{t}}\,t^{\sigma^{2}-\theta_{0}^{2}-\theta_{t}^{2}}\tau(\bm{\theta};t).
\end{equation}
This new $\tau$-function changes the Hamiltonian, but does not change
the equations of motion, as it is multiplied by a pure function of
$t$. The complete expansion of the Painlevé VI $\tau$-function,
adapted to our definition \eqref{eq:66}, is given by
\begin{equation}
  \tau(\bm{\theta};t)
  =\sum_{n\in\mathbb{Z}}C(\bm{\theta},\sigma+n)
  s^{n} t^{n(n+2 \sigma)}{\cal
    B}(\bm{\theta},\sigma+n;t),
\label{eq:taufunctionexpansion}
\end{equation}
where we assume that the real part of sigma $\Re \sigma$ obeys
\begin{equation}
  \label{eq:87}
  0 \leq \Re \sigma < \frac{1}{2}.
\end{equation}
The structure constants are given in terms of Barnes
functions\footnote{Defined by the functional relation
  $G(z+1)=\Gamma(z)G(z)$, with $\Gamma(z)$ being the Euler gamma
  function. For further properties, see appendix A of
  \cite{Gamayun:2013auu}.}
\begin{equation}
\label{eq:structconsts}
C(\bm{\theta},\sigma)=\frac{\prod_{\epsilon,\epsilon'=\pm}G(1+\theta_t+
  \epsilon\theta_0+\epsilon'\sigma)G(1+\theta_1+\epsilon\theta_\infty+
  \epsilon'\sigma)}{\prod_{\epsilon=\pm}G(1+2\epsilon\sigma)},
\end{equation}
and the ${\cal B}$'s are the $c=1$ conformal blocks, given by the
AGT combinatorial series
\begin{equation}
\label{eq:cblockstau}
{\cal B}(\bm{\theta},\sigma;t)=
\sum_{
  \lambda,\mu\in\mathbb{Y}}{\cal
  B}_{\lambda,\mu}(\bm{\theta},\sigma)t^{|\lambda|+|\mu|}, 
\end{equation}
summing over pairs of Young diagrams $\lambda,\mu$ with
\begin{multline}
\label{eq:cblockstaucoeffs}
{\cal
  B}_{\lambda,\mu}(\bm{\theta},\sigma)=\prod_{(i,j)\in\lambda}
\frac{((\theta_t+\sigma+i-j)^2-\theta_0^2)((\theta_1+\sigma+i-j)^2 
-\theta_\infty^2)}{ h_\lambda^2(i,j)(\lambda'_j+\mu_i-i-j+1+2\sigma)^2}
\times \\
\prod_{(i,j)\in\mu}
\frac{((\theta_t-\sigma+i-j)^2-\theta_0^2)((\theta_1-\sigma+i-j)^2 
-\theta_\infty^2)}{ h_\mu^2(i,j)(\lambda_i+\mu'_j-i-j+1-2\sigma)^2},
\end{multline}
where $(i,j)$ denotes the box in the Young diagram $\lambda$,
$\lambda_i$ the number of boxes in row $i$, $\lambda'_j$ the number of
boxes in column $j$ and $h_{\lambda}(i,j)=\lambda_i+\lambda'_j-i-j+1$
its hook length. A proof for this expansion is available in
\cite{Iorgov2015} and alternatively in \cite{Bershtein:2014yia}.

\subsection{Solving the condition $\lambda(x)=x$}
\label{sec:solving-lambdax=x}

Inspired by \cite{Litvinov:2013sxa}, our approach to find $H_{x}$ is
to solve the condition $\lambda(x) = x$ by assuming a series expansion
for $s=s(\sigma,x)$ in $x$. The Painlevé property implies that the PVI
solution is regular at $t=x$ for any $x\neq 0,1,\infty$
\cite{Litvinov:2013sxa,Guzzetti2012b,Iwasaki:1991}. As we are going to
see below, this fact is consistent with the $\tau$-function also being
regular at $t=x$.
Okamoto's formula \cite{Okamoto1986} gives $\lambda(t)$ in terms of a
ratio of $\tau$-functions
\begin{equation}
  \label{eq:okamoto}
  \lambda(t)-t = \frac{t(t-1)}{2\theta_{\infty}}\Dfrac{}{t}\log\frac{\tau(\bm{\theta}_{0,-};t)}{\tau(\bm{\theta}_{0,+};t)},
\end{equation}
where $\tau(\bm{\theta},t)$ is given by
\eqref{eq:taufunctionexpansion}. Evaluated at $t=x$, the left hand side
vanishes because of our initial condition. Assuming that
$\theta_{\infty}\neq 0$ and neither of the $\tau$-functions above
vanishes or blows up at $\lambda(x)=x$, we have that
\begin{equation}
  \label{eq:taucondition} 
  \left[
    \frac{d}{dt}\left[\tau\left(\bm{\theta}_{0,-};t\right)\right]\tau\left(\bm{\theta}_{0,+};t\right)-\frac{d}{dt}\left[\tau\left(\bm{\theta}_{0,+};t\right)\right]\tau\left(\bm{\theta}_{0,-};t\right)
  \right]\Bigg\lvert_{t=x}  = 0.
\end{equation}

In the following, we rewrite the condition \eqref{eq:taucondition} as
a double series expansion. To do this, let us introduce some
definitions.  The isomonodromic Hamiltonian \eqref{eq:8} and the
accessory parameter expansion \eqref{eq:litvinov_cond2} do not depend
on the normalization of the $\tau$-function, so we normalize the
structure constants as
\begin{equation}
  \label{eq:70}
  \bar{C}_{n}(\btheta,\sigma) \defeq \frac{C(\btheta,\sigma+n)}{C(\btheta,\sigma)}.
\end{equation}
As we prove in Appendix \ref{sec:painleve-vi-tau}, these ratios can be
factorized as
\begin{equation}
  \label{eq:cbar}
  \cbar = \mathcal{C}_{n}(\btheta,\sigma) A(\btheta,\sigma)^{n},\quad (n>0)
\end{equation}
where $\mathcal{C}_{n}(\btheta,\sigma)$ is given by \eqref{eq:14} and
$A(\btheta,\sigma)$ is given by \eqref{eq:10}. Another important
formula is
\begin{equation}
  \label{eq:72}
  \bar{C}_{n}(\bm{\theta}_{0,-}\,,\sigma) =
  f_{n}(\bm{\theta}\,,\sigma)
  \bar{C}_{n}(\bm{\theta}_{0,+}\,,\sigma),\quad (n>0)
\end{equation}
where
\begin{equation}
  \label{eq:68}
  f_{n}(\bm{\theta},\sigma) = \prod_{k=1}^{|n|}\frac{
    \left(
      \sigma+k-\tfrac12-\theta_{\infty}
    \right)^{2}-\theta_{1}^{2}}{
    \left(
      \sigma+k -\tfrac12+ \theta_{\infty}
    \right)^{2}-\theta_{1}^{2}},\quad f_{0}(\bm{\theta},\sigma)=1.
\end{equation}
For $n<0$, we change $\sigma \rightarrow -\sigma$ in the formulas
above. 

Plugging \eqref{eq:70} and \eqref{eq:cbar} into
\eqref{eq:taufunctionexpansion}, we get
\begin{equation}
  \label{eq:71}
  \tau(\bm{\theta};t)=\sum_{n\in\mathbb{Z}} \mathcal{C}_{n}(\btheta,\sigma)
  {\cal
    B}(\bm{\theta},\sigma+n;t) t^{n^{2}} [X(\btheta,\sigma,s;t)]^{n},
\end{equation}
where
\begin{equation}
  \label{eq:89}
  X(\btheta,\sigma,s;t) \defeq A(\btheta,\sigma) s\,
t^{2\sigma}.
\end{equation}
We can rewrite the $c=1$ blocks \eqref{eq:cblockstau} in terms of levels $L$
\begin{equation}
  \label{eq:73}
  {\cal B}(\bm{\theta},\sigma+n;t)=
\sum_{L=0}^{\infty}
  {\cal
  B}_{n}^{(L)}(\btheta,\sigma)\,t^{L}, 
\end{equation}
where
\begin{equation}
  \label{eq:67}
  \mathcal{B}_{n}^{(L)} (\btheta,\sigma)= \sum_{\left|\lambda\right|+\left|\mu\right|=L}
\mathcal{B}_{\lambda,\mu}(\bm{\theta},\sigma+n)
\end{equation}
is a restricted sum at level $L$ of the coefficients
\eqref{eq:cblockstaucoeffs}.  We will suppress the monodromy arguments
in what follows, for simplicity, except the ones that are
shifted. Using the definitions \eqref{eq:71} and \eqref{eq:73}, we get
\begin{align}
\label{eq:tauX}
\tau\left(t\right) & = 
                         \sum_{n\in\mathbb{Z}}\mathcal{C}_{n}\sum_{L=0}^{\infty}\mathcal{B}_{n}^{(L)}\,[X(t)]^{n}\, t^{n^{2}+L},\\[5pt]
\frac{d}{dt}\tau\left(t\right) & = 
                                     \sum_{n\in\mathbb{Z}}\mathcal{C}_{n}\sum_{L=0}^{\infty}\left(n\left(n+2\sigma\right)+L\right)\mathcal{B}_{n}^{(L)}\,[X(t)]^{n}\, t^{n^2+L-1}\label{eq:derivtauX},
\end{align}
where we used
\begin{equation}
  \label{eq:80}
  \Dfrac{X(t)}{t} = \frac{2\sigma}{t}X(t).
\end{equation}
Using \eqref{eq:tauX} and \eqref{eq:derivtauX} in
\eqref{eq:taucondition}, with the properly shifted monodromies and
considering \eqref{eq:72}, the equation \eqref{eq:taucondition} can be
rewritten as
\begin{equation}
  \label{eq:3}
  \sum_{n,p\in\mathbb{Z}}\sum_{L,M=0}^{\infty}{ D}\begin{bmatrix}
                                                              L & M \\
                                                              n & p
\end{bmatrix}x^{n^{2}+p^{2}+L+M-1}[X(x)]^{n+p}=0,
\end{equation}
where
\begin{multline}
\label{eq:dcoeffs}
D
\begin{bmatrix}
L & M\\
n & p
\end{bmatrix}
\equiv
\mathcal{C}_{n}(\btheta_{0,+},\sigma)\,\mathcal{C}_{p}(\btheta_{0,+},\sigma)\left[f_{n}(\btheta_{0,+},\sigma)\,\mathcal{B}_{n}^{(L)}\left(\bm{\theta}_{0,-},\sigma
  \right)
  \mathcal{B}_{p}^{(M)}\left(\bm{\theta}_{0,+},\sigma\right)\right.
-\\-
\left.  f_{p}(\btheta_{0,+},\sigma)\, \mathcal{B}_{n}^{(L)}
  \left(\bm{\theta}_{0,+},\sigma\right) \mathcal{B}_{p}^{(M)}
  \left(\bm{\theta}_{0,-},\sigma\right)\right]
\times\left[n\left(n+2\sigma\right)+L\right].
\end{multline}
Given that $\lambda(t)$ admits a regular expansion near $t=x$, so does
the right hand side of \eqref{eq:okamoto}. This implies that $X(x)$
also admits a  Taylor expansion for $x$ small
\begin{equation}
  \label{eq:74}
  X(\bm{\theta},\sigma,s(\sigma,x);x) = \sum_{k=0}^{\infty}X_{k}(\bm{\theta},\sigma)x^{k},
\end{equation}
where we assume that $s=s(\sigma,x)$ is the source of the series
expansion. Our goal is to extract the coefficients
$X_{k}=X_{k}(\bm{\theta},\sigma)$ by solving \eqref{eq:3} order by
order in $x$.  Therefore, this procedure gives us the coefficients of
the $s=s(\sigma,x)$ series solution.

We rewrite \eqref{eq:3} as a double series condition
\begin{equation}
  \label{eq:84}
  \sum_{\alpha=0}^{\infty}\sum_{\beta = -n_{\alpha}}^{n_{\alpha}}e_{\alpha,\beta}\,x^{\alpha}[X(x)]^{\beta} = 0,
\end{equation}
with
\begin{equation}
  \label{eq:85}
  e_{\alpha,\beta}=\sideset{}{'}\sum_{n,p,L,M} D\left[\begin{array}{cc}
L & M \\
n & p
\end{array}\right],
\end{equation}
where the $\sum^{'}$ means that the summation is over all
$n,p\in\mathbb{Z}$; $L,M=0,1,2,\dots$ such that
$n^{2}+p^{2}+L+M-1=\alpha$ and $n+p=\beta$. From \eqref{eq:dcoeffs}, we see that
\begin{equation}
  \label{eq:90}
  D
\begin{bmatrix}
0 & M\\
0 & p
\end{bmatrix} 
=0,\quad \forall\, p,M.
\end{equation}
 This means that the $\alpha =-1$ term does not
contribute to \eqref{eq:3}. For fixed $\alpha$, we have that
$-n_{\alpha} \leq \beta \leq n_{\alpha}$, where
$n_{\alpha}=\left\lfloor \sqrt{\alpha+1}\right\rfloor $, with
$\left\lfloor x\right\rfloor $ being the floor function of $x$. To
explain the $\beta$ constraint, let us consider both $n$ and $p$ to be
positive (or both negative). For fixed $\alpha$, the upper bound on
$\beta$ can be reached when $L=M=0$ so that
$\alpha+1=n^{2}+p^{2}\geq\left(n+p\right)^{2}=\beta^{2}$.

Let us now go back to \eqref{eq:3}. For $\alpha=0$, we have
$-1\leq \beta \leq 1$ and thus
\[
D\begin{bmatrix}
0 & 0\\
-1 & 0
\end{bmatrix}X_{0}^{-1}+D\begin{bmatrix}
1 & 0\\
0 & 0
\end{bmatrix}+D\begin{bmatrix}
0 & 0\\
1 & 0
 \end{bmatrix}X_{0}=0,
\]
which, using \eqref{eq:dcoeffs}, \eqref{eq:67}, \eqref{eq:68} and
\eqref{eq:14}, we can show it is equivalent to
\begin{equation}
  \label{eq:17}
  \frac{(\theta_{t}+\sigma)^{2}-\theta_{0}^{2}}{4\sigma^{2}}X_{0}^{-1}+ \left( \frac{\theta_{0}^{2}-\sigma^{2}-\theta_{t}^{2}}{2\sigma^{2}}\right) + \frac{(\theta_{t}-\sigma)^{2}-\theta_{0}^{2}}{4\sigma^{2}}X_{0}= 0.
\end{equation}
This is essentially the same equation presented in
\cite{Litvinov:2013sxa} obtained from the PVI solution\footnote{With
  the corresponding Bäcklund transformation
  $\theta_{0}\rightarrow \theta_{1}$ and
  $\theta_{t} \rightarrow \theta_{\infty}+\tfrac12$.}. It gives two
solutions for $X_{0}$ and we choose $X_{0}=1$ at this order, for
consistency with the asymptotics of the PVI solution, as discussed in
\cite{Litvinov:2013sxa}. Plugging back $X(x) = 1 + X_{1}x+\ldots$ to
\eqref{eq:84}, the higher order equations in $x$ will give only one
solution for the coefficients $X_{k}$. The next two coefficients are
given in Appendix \ref{sec:numer-checks-access} in terms of $\sigma$
and $\delta$'s.

\subsection{Accessory Parameter Expansion}
\label{sec:access-param-expans}

Substituting the definition \eqref{eq:66} into
\eqref{eq:litvinov_cond2} and then into \eqref{eq:htok}, we get
$H_{x}$ in terms of the $\tau$-function
\begin{align}
  \label{eq:normaltocanonical}
  H_{x}&=\frac{\delta_{0}+\delta_{x}-\delta_{\sigma}}{x}+\frac{(1-2\theta_{1})(1-2\theta_{x})}{2(x-1)}+\Dfrac{}{t}\log\tau (\bm{\theta}_{0,+};t) \Big\rvert_{t=x}  .
\end{align}
Notice that the first term in \eqref{eq:normaltocanonical} is already
the expected answer from CFT, as it comes from the leading behavior of
the classical conformal block. The final step in our algorithm is to
take the series solution for $X(x)$ obtained above and plug it back
into the $\tau$-function in \eqref{eq:normaltocanonical}, with the
substitution $\theta_{t} =\theta_{x}-\frac12 $. This computation is
straightforward but demanding, so we will not discuss it here in
general form.  We will show below how the computation works for the
first non-trivial term. This will make it clear how to proceed to
obtain higher-order terms.

Remembering that $X(t) = A s t^{2\sigma}$, under the assumption
$0< \Re \sigma < \tfrac12$, we have the small $t$ expansion of \eqref{eq:71}
\begin{align*}
  \label{eq:75}
  \tau(t) &= 1 + 
  \left(
    \mathcal{C}_{1}X(t)+ \mathcal{C}_{-1}X(t)^{-1}+\mathcal{B}_{0}^{(1)}
  \right) t +\mathcal{O}(t^{2(1\pm \Re\sigma)}
  ),
\end{align*}
and
\begin{equation}
  \label{eq:79}
  \Dfrac{}{t}\tau(t) =(1-2\sigma)\mathcal{C}_{-1}X(t)^{-1} +(1+2\sigma) \mathcal{C}_{1}X(t)+
  \mathcal{B}_{0}^{(1)} +\mathcal{O}(t^{1\pm 2\Re\sigma}
  ),
\end{equation}
The $\tau$-function has no well-defined Taylor expansion around $t=0$
because its first derivative diverges as $t^{-2\sigma}$. On the other
hand, it is possible to do such expansion near a regular point $t=x$,
for $x$ sufficiently close to zero. In this sense, we are allowed to
write
\begin{equation}
  \label{eq:78}
  \Dfrac{}{t}\log\tau(t) \sim 
  \frac{(1-2\sigma)\mathcal{C}_{-1}X(t)^{-1} +(1+2\sigma) \mathcal{C}_{1}X(t)+
    \mathcal{B}_{0}^{(1)} }{1 + 
    \left(
      \mathcal{C}_{1}X(t)+ \mathcal{C}_{-1}X(t)^{-1}+\mathcal{B}_{0}^{(1)}
    \right) t}.
\end{equation}
Using that $X(t=x)=1+X_{1}x+\ldots$ and
$\mathcal{B}_{0}^{(1)}= \mathcal{B}_{0,1}+\mathcal{B}_{1,0}$, we have
\begin{equation}
  \label{eq:76}
  \Dfrac{}{t}\log\tau(t)\bigg\rvert_{t=x} = (1-2\sigma)\mathcal{C}_{-1} +(1+2\sigma) \mathcal{C}_{1}+
  \mathcal{B}_{0,1}+\mathcal{B}_{1,0}+ \mathcal{O}(x)
\end{equation}
to leading order in $x$. From \eqref{eq:14}, we see that
\begin{multline}
  \label{eq:77}
 (1+ 2\sigma) \mathcal{C}_{ 1} +(1- 2\sigma) \mathcal{C}_{-1}
 =\\[5pt]=
 \frac{\left[\left(\theta_{t}-\sigma\right)^{2}-\theta_{0}^{2}\right]
\left[
  \left( \theta_{\infty}+\sigma\right)^{2}-\theta_{1}^{2}\right]}{4\sigma^{2}(1+2\sigma)}+
\frac{\left[\left(\theta_{t}+\sigma\right)^{2}-\theta_{0}^{2}\right]\left[
    \left(
      \theta_{\infty}-\sigma
    \right)^{2}-\theta_{1}^{2}\right]}{4\sigma^{2}(1-2\sigma)}
\end{multline}
and from \eqref{eq:cblockstaucoeffs}
\begin{multline}
  \label{eq:81}
\mathcal{B}_{1,0}+\mathcal{B}_{0,1} =\\[5pt]
= \frac{\left[\left(\theta _t-\sigma \right)^2-\theta_0^2\right]
\left[\left(\theta
      _1-\sigma \right)^{2}-\theta_{\infty }^2\right]}{4\sigma ^2}
+\frac{\left[\left(\theta_t+\sigma\right)^2-\theta _0^2\right]\left[\left(\theta _1+\sigma \right)^2-\theta_{\infty }^2\right] }{4 \sigma ^2}.
\end{multline}
If we carefully sum \eqref{eq:77} and \eqref{eq:81}, substituting
$\theta_{i}^{2}=1/4-\delta_{i}$ and the shifted monodromy values
$\theta_{t} \rightarrow \theta_{x}-\frac12, \theta_{\infty}\rightarrow
\theta_{\infty}+\frac12$, we get
\begin{equation}
  \label{eq:82}
\frac{(\delta_{\sigma}+\delta_{x}-\delta_{0})(\delta_{\sigma}+\delta_{1}-\delta_{\infty})}{2\delta_{\sigma}}
+(1-2\theta_{1})(1-2\theta_{x}).
\end{equation}
The second term above cancels with the second term of
\eqref{eq:normaltocanonical} and the first term above matches the
standard CFT result \cite{Litvinov:2013sxa}. The next order
calculation is much more complicated, so we show here the result for
\eqref{eq:normaltocanonical} only up to first order in $x$
\begin{equation}
  \label{eq:26}
  H_{x} = \frac{\delta_{0}+\delta_{x}-\delta_{\sigma}}{x} +
  \frac{(\delta_{0}-\delta_{\sigma}-\delta_{x})(\delta_{\sigma}+\delta_{1}-\delta_{\infty})}{2\delta_{\sigma}}
  +\mathcal{O}(x).
\end{equation}
We present the next two terms in the expansion above in appendix
\ref{sec:numer-checks-access}. 

We implemented the algorithm of this section in a computer algebra
program, as we do not have the explicit series solution for the
constraint \eqref{eq:84}. We tested the accessory parameter expansion
up to order $x^{2}$ analytically by comparing with the semiclassical
limit of the CFT conformal block \cite{Ferrari2012a,Litvinov:2013sxa}
using the inverse gram matrix approach
\cite{Mironov:2010zs,Gamayun:2013auu}.  We also checked this algorithm
numerically, substituting the $\theta$'s with 
some fixed numbers from the beginning, up to order $x^{5}$. More
details are given in appendix \ref{sec:numer-checks-access}. This
gives strong evidence that the isomonodromic approach presented here
reproduces the classical conformal block expansion. Although we do not
have a proof to all orders in $x$, we believe that our discussion
above on the semiclassical limit and isomonodromic deformations, given
the exponentiation hypothesis and the assumptions on the conformal
weights, is enough mathematical evidence for the $\tau$-function
expansion.

In addition, according to our tests, it is numerically faster to
calculate the accessory parameter expansion with our approach than
inverting the Gram matrix and taking the semiclassical limit. We did
not compare our approach with the Zamolodchikov recurrence formula,
but, then again, the exact large $c$ limit is also demanding in this
case, while our algorithm already gives the analytic coefficients in
the semiclassical limit.

\section{Isomonodromic Approach to Classical Conformal Blocks}
\label{sec:corr-from-access}

Now that we found the accessory parameter using the isomonodromic
$\tau$-function, an interesting question is if we can find an
analogous formula for classical conformal blocks. One way to approach
this problem is to use the symplectic structure of isomonodromic
deformations \cite{Litvinov:2013sxa,Novaes2014c}. As we discussed
above, $S_{\sigma}(\lambda,t)$ is the action of the isomonodromic
Hamiltonian system
\begin{equation}
  \label{eq:11}
  \dot{\lambda} = \Dpfrac{H}{p},\quad \dot{p} = -\Dpfrac{H}{\lambda},
\end{equation}
determined by the Hamilton--Jacobi equation
\eqref{eq:taufunctioneq}. The associated symplectic structure is given
by
\begin{equation}
  \label{eq:symplecticstruct}
  \Omega = dp\wedge d\lambda - dH\wedge dt.
\end{equation}
Defining the one-form
\begin{equation}
  \label{eq:5}
  \omega  = pd\lambda - H dt,
\end{equation}
we recover $\Omega = d\omega$. We can define the action as the
generating function of the canonical transformation from $(p,\lambda)$
to action-angle coordinates $(\sigma,
\nu)$
\begin{equation}
  \label{eq:action}
  dS_{\sigma} = pd\lambda - Hdt + 
  \nu d\sigma. 
\end{equation}
Here $\sigma=\sigma_{0t}$ and 
$\nu$ is the canonically conjugate variable to $\sigma$
\cite{Litvinov:2013sxa,Nekrasov:2011bc}. These coordinates
parameterize the moduli space of $\slc$ flat-connections, similarly to
$(\sigma,s)$.  The transformation
\begin{align}
  \label{eq:canonicaltransf}
 p(\mu) &= \mu + \sum_{i=0,1,t}\frac{1-2\theta_{i}}{2(\lambda-a_{i})},\\[5pt]
  H(\btheta;\lambda,p(\mu),t) &= K(\btheta;\lambda,\mu,t) +\nonumber
  \\[5pt] &
+\frac{(1-2\theta_{0})(1-2\theta_{t})}{2t} + \frac{(1-2\theta_{1})(1-2\theta_{t})}{2(t-1)} +\frac{1-2\theta_{t}}{2(\lambda-t)}
\end{align}
between the normal form \eqref{eq:47} and the canonical form
\eqref{eq:garnier} is canonical with respect to $\Omega$
\cite{Iwasaki:1991}. This induces a transformation
\begin{equation}
  \label{eq:91}
  S_{\sigma} = S_{\sigma}^{c} + g(\lambda,t)
\end{equation}
by the function
\begin{equation}
  \label{eq:12}
  g(\lambda,t) = \log
  \left[
    \prod_{i=0,1}(t-a_{i})^{-2(\frac12-\theta_{i})(\frac12-\theta_{t})}\prod_{i=0,1,t}(\lambda-a_{i})^{\frac12-\theta_{i}}
  \right]
\end{equation}
and thus
\begin{equation}
  \label{eq:action}
  dS^{c}_{\sigma} = \mu d\lambda - Kdt + \nu d\sigma. 
\end{equation}

Integrating \eqref{eq:action} over a solution of the isomonodromic
system, as the action-angle variables are constant on the orbits, we get
\begin{equation}
  \label{eq:28}
  S^{c}_{\sigma}(\lambda(t),t) 
  = \int^{(\lambda(t),t)}_{(\lambda(0),0)}
  \left(
    \mu d\lambda' - Kdt'
  \right) = \int^{t}_{0}
  \left(
    \mu\Dfrac{\lambda'}{t'}-K
  \right)dt'.
\end{equation}
Let us try to simplify the PVI Lagrangian as much as possible. First,
we notice that
\begin{equation}
  \label{eq:29}
  \dot{\lambda} = \Dpfrac{K}{\mu} =  \frac{\lambda(\lambda-1)(\lambda-t)}{t(t-1)}
  \left[2\mu-\left(\frac{2\theta_0}{\lambda}+\frac{2\theta_1}{\lambda-1}+
      \frac{2\theta_t-1}{\lambda-t}\right)\right]
\end{equation}
and therefore
\begin{equation}
  \label{eq:51}
\mu\dot{\lambda} - K =  \frac{\lambda(\lambda-1)(\lambda-t)}{t(t-1)}\mu^{2} - \frac{\kappa (\lambda-t)}{t(t-1)}.
\end{equation}
Using Okamoto's formula \eqref{eq:okamoto}, we can integrate the second term in \eqref{eq:51}
\begin{equation}
  \label{eq:52}
    S^{c}_{\sigma}(\lambda(t),t) = \int_{0}^{t}  \frac{\lambda(t')(\lambda(t')-1)(\lambda(t')-t')}{t'(t'-1)}\mu(t')^{2}\, dt' -\frac{\kappa}{2\theta_{\infty}}\log\frac{\tau(\btheta_{0,-};t)}{\tau(\btheta_{0,+};t)}\Bigg\lvert_{0}^{t}.
\end{equation}
As both $\tau$-functions have the same leading behaviour as $t$ goes
to zero, only the upper limit contributes to the second term
above. Another equation proved by Okamoto in \cite{Okamoto1986} is
\begin{align}
  \label{eq:53}
  \lambda(\lambda-1)\mu &= -\kappa_{1} (\lambda-t) + t(t-1)\Dfrac{}{t}
                          \log\left[   \frac{\tau(\btheta_{-,0};t)}{\tau(\btheta_{0,+};t)}t^{\theta_{0}+\theta_{t}-\tfrac14}
                          \right]\nonumber\\[10pt]
                        &= \frac{t(t-1)}{2\theta_{\infty}}\Dfrac{}{t}\log
                          \left[
                          \left(
                          \frac{\tau_{+}}{\tau_{-}}
                          \right)^{\kappa_{1}}
                          \left(t^{\theta_{0}+\theta_{t}-\tfrac14}
                          \frac{\tau_{0}}{\tau_{+}}
                          \right)^{2\theta_{\infty}}
                          \right],
\end{align}
where  we used \eqref{eq:okamoto} in the second line and
\begin{equation}
  \label{eq:54}
  \tau_{+} \defeq \tau(\btheta_{0,+};t),\quad \tau_{-}\defeq \tau(\btheta_{0,-};t),\quad \tau_{0}\defeq \tau(\btheta_{-,0};t).
\end{equation}
It also follows from \eqref{eq:okamoto} that
\begin{align}
  \label{eq:63}
  \lambda = \frac{t(t-1)}{2\theta_{\infty}}\Dfrac{}{t}\log
  \left[
  (t-1)^{2\theta_{\infty}}\frac{\tau_{-}}{\tau_{+}}
  \right],\quad   \lambda -1= \frac{t(t-1)}{2\theta_{\infty}}\Dfrac{}{t}\log
  \left[
  t^{2\theta_{\infty}}\frac{\tau_{-}}{\tau_{+}}
  \right].
\end{align}
We then use \eqref{eq:53} and \eqref{eq:63} to express the action only
in terms of $\tau$-functions
\begin{equation}
  \label{eq:action_final}
  S^{c}_{\sigma}(\lambda(t),t) = \frac{1}{2\theta_{\infty}}\int_{0}^{t}I[\tau_{\pm},\tau_{0};t'] \, dt' -\frac{\kappa}{2\theta_{\infty}}\log
  \left(
    \frac{\tau_{-}}{\tau_{+}}
  \right),
\end{equation}
where
\begin{equation}
  \label{eq:64}
  I[\tau_{\pm},\tau_{0};t]=   \frac{\Dfrac{}{t}\log\left(\frac{\tau_{-}}{\tau_{+}}\right)
    \left\{
      \Dfrac{}{t}\log
      \left[
        \left(
          \frac{\tau_{+}}{\tau_{-}}
        \right)^{\kappa_{1}}
        \left(t^{\theta_{0}+\theta_{t}-\tfrac14}
          \frac{\tau_{0}}{\tau_{+}}
        \right)^{2\theta_{\infty}}
      \right]
    \right\}^{2}}{\Dfrac{}{t}\log\left(t^{2\theta_{\infty}}\frac{\tau_{-}}{\tau_{+}}\right)\Dfrac{}{t}\log\left((t-1)^{2\theta_{\infty}}\frac{\tau_{-}}{\tau_{+}}\right)}.
\end{equation}
Although \eqref{eq:action_final} is still complicated, substitution of
\eqref{eq:71} into \eqref{eq:64} should give a formula for it only in
terms of known functions of the monodromies. However, to get a closed
form for the 4-point classical conformal block \eqref{eq:intro1}, we
still need to impose the condition $\lambda(x)=x$ into the resulting
formula, following the procedure of the previous section. As we can
only solve this condition order-by-order in $x$, a closed form is
still out of reach. But, as we are going to see below, some special
cases can be 
tractable. We leave the detailed study of this formula for future
work.

\subsection{Recovering the 4-point Classical Conformal Block}
\label{sec:recovering-4-point}

The 5-point semiclassical block \eqref{eq:sclimit5pt} is given by
\begin{multline}
  \label{eq:sc5ptfromtau}
  \langle V_{\delta_{0}}(0)V_{\delta_{t}}(t)
  \,\Pi_{\substack{\sigma}}\,
  \varphi_{H}(\lambda(t))\Pi_{\substack{\sigma \pm \frac{1}{2}}}
  V_{\delta_{1}}(1)V_{\delta_{\infty}}(\infty) \rangle \sim \exp{
    \left( \frac{1}{b^{2}}S_{\sigma}^{\pm}(\lambda(t),t
      )   \right)}  \\[5pt]
  = (\lambda(t)-t)^{(\frac12-\theta_{t})/b^{2}}
  \prod_{i=0,1}\left[(t-a_{i})^{-2(\frac12-\theta_{i})(\frac12-\theta_{t})/b^{2}}(\lambda-a_{i})^{(\frac12-\theta_{i})/b^{2}}\right]
  \exp{ \left( \frac{1}{b^{2}} S_{\sigma}^{c,\pm}(\lambda(t),t)
    \right)},
\end{multline}
where we used \eqref{eq:91}. We label the fields by their classical
weights $\delta_{i}$ and $\Pi_{\sigma}$ corresponds to the projection
operator onto the intermediate state with momentum $P=i \sigma/b$.
Notice that because $\lambda(t)$ is a solution of isomonodromic
deformations, we can change the position of the heavy degenerate field
without changing the other monodromies. This has a nice $AdS_{3}$
interpretation, as degenerate insertions are conical defects in the
bulk \cite{Witten1988,Raeymaekers2015,Hulik2017}. The PVI action
governs the evolution of this conical defect in a way that the
monodromies do not change. Therefore, we impose the boundary condition
for the isomonodromic flow at $t=x$ to be $\lambda(x) = x$. This
entails to taking the fusion
\begin{equation}
  \label{eq:fusion}
  \varphi_{H}(\lambda(t))V_{\delta_{t}}(t) = C_{+} (\lambda(t)-t)^{(\frac12-\theta_{t})/b^{2}}V_{\delta(\frac{1}{2}+\theta_{t})}(t) + C_{-} (\lambda(t)-t)^{(\frac12+\theta_{t})/b^{2}}V_{\delta(\frac12-\theta_{t})}(t) .
\end{equation}
Our choice \eqref{eq:sc5ptfromtau} clearly corresponds to taking
$C_{-} = 0$ so that the leading term in the OPE cancels with the
appropriate term in right hand side of
\eqref{eq:sc5ptfromtau}. Therefore, we end up with
\begin{equation}
  \label{eq:sc4ptfromtau}
  \langle
  V_{\delta_{0}}(0)V_{\delta_{x}}(x)\Pi_{\nu}V_{\delta_{1}}(1)V_{\delta_{\infty}}(\infty) 
  \rangle  =
\prod_{i=0,1}(x-a_{i})^{-2(\frac12-\theta_{i})(\frac12-\theta_{x} )/b^{2}} 
\exp{
  \left(
  \frac{1}{b^{2}}S_{\sigma}^{c,\pm}(x;\,\btheta)  \right)} ,
\end{equation}
where we defined $\theta_{t}= \theta_{x}-\frac12$ and
$\nu = \sigma \pm \frac12$.  The identification of the classical
conformal block with the $\tau$-function in \eqref{eq:action_final} is
a new result with important technical consequences, as the
$\tau$-function is a linear combination of $c=1$ conformal blocks, as
described in \cite{Gamayun2012}.

\subsubsection*{Linear Dilaton Case}
\label{sec:free-boson-case}

In some special cases, the PVI solutions dramatically simplify. The
simplest example is obtained by assuming
\begin{equation}
  \label{eq:57}
  \sum_{i=0,1,t}\theta_{i} +
  \theta_{\infty} = \frac{1}{2}
\end{equation}
and $\sigma_{ij} = \theta_{i}+\theta_{j}$, $i,j=0,1,t$. This implies
that $\kappa_{2}+1 = 0$ (see \eqref{eq:20}) and thus $\mu(t)=0$ is
consistent with the equations of motion \eqref{eq:18}
\cite{Iwasaki:1991}. Therefore, \eqref{eq:52} gives $S_{\sigma}^{c}=0$
and the PVI action is given by
\begin{equation}
  \label{eq:58}
  S_{\sigma}(\lambda(t),t) = \log
  \left[
    (t(t-1))^{\theta_{t}-\frac12}\prod_{i=0,1,t}(\lambda(t)-a_{i})^{\frac12-\theta_{i}}
  \right]
\end{equation}
leading to the 4-point conformal block \eqref{eq:sc4ptfromtau} 
\begin{equation}
  \label{eq:freeboson}
  \langle
  V_{\delta_{0}}(0)V_{\delta_{x}}(x)\Pi_{\nu}V_{\delta_{1}}(1)V_{\delta_{\infty}}(\infty) 
  \rangle  = x^{-2\alpha_{0}\alpha_{x}}(x-1)^{-2\alpha_{1}\alpha_{x}},
\end{equation}
where $\nu = \theta_{0}+\theta_{x}$ and
\begin{equation}
  \label{eq:59}
  \alpha_{i} = 
  \left(
    \frac{1}{2} -\theta_{i}
  \right)\frac{1}{b^{2}},
\end{equation}
according to the alternative Liouville definition
$\Delta_{i} = \alpha_{i}(Q-\alpha_{i})$.  This corresponds to the
semiclassical limit of the linear dilaton correlator obeying the
screening condition \cite{Ribault:2014aa}
\begin{equation}
  \label{eq:60}
  \sum_{i=0,1,x,\infty}\alpha_{i} = Q \rightarrow \frac{1}{b^{2}}\quad \text{as} \quad b\rightarrow 0.
\end{equation}
Notice that, in the $c=1$ interpretation, the $\tau$-function is a
hypergeometric function \cite{Gamayun:2013auu,iorgov2013painleve},
showing the non-triviality of this case in comparison to the
$c=\infty$ result.

\section{Conclusions}
\label{sec:discussion}

In this work, we discussed the deep mathematical relation between the
large central charge limit of conformal blocks and the isomonodromic
$\tau$-function. We recovered the accessory parameter
\eqref{eq:normaltocanonical} of the 4-point Fuchsian equation
\eqref{eq:heun} using the isomonodromic $\tau$-function, with the
additional fusion constraint $\lambda(x)=x$.  We believe our approach
gives a more straightforward algorithm than the one presented in
\cite{Litvinov:2013sxa}.

The isomonodromic approach is relevant to applications of the theory
of differential equations, as there is no need to take any
semiclassical limit, in comparison to the CFT approach to calculate
the accessory parameters.  Thus, it can be applied to any particular
problem that is governed by a Heun's equation.  Recently, Piatek and
Pietrykowski \cite{Piatek2017} recovered Floquet solutions of Heun's
equation using CFT. This result nicely complements our discussion of
accessory parameters. These techniques 
might be used in many concrete
applications. In particular, the isomonodromic method has already
proved useful for scattering problems in black hole physics
\cite{Novaes2014c,daCunha:2015fna,Amado:2017kao}.

We can also use it to calculate accessory parameters of confluent
cases of Heun's equation, using the corresponding $\tau$-functions
described in \cite{Gamayun:2013auu,Bonelli2016}. Those are connected
to the so-called irregular conformal blocks
\cite{Gaiotto2013,Piatek2014,Nagoya2015}, which deserve to be better
understood in CFT applications. Finally, we notice that it is
straightforward to generalize the isomonodromic method for Fuchsian
equations with any number of singular points by using the appropriate
$\tau$-function \cite{Iorgov2015,Gavrylenko2016}. Our algorithm can
then be generalized to find the accessory parameters and classical
conformal blocks of $n$-point correlators.

The integral formula \eqref{eq:action_final} for the PVI action is the
first step to a closed expression for the 4-point classical conformal
block \eqref{eq:intro1}. Using the $\tau$-function $c=1$ expansion, it
should be possible to fully integrate the PVI action and impose the
fusion constraint to obtain \eqref{eq:intro1}. This has many potential
applications, in particular, related to the AdS/CFT correspondence:
the emergence of $AdS_{3}$ gravity backgrounds
\cite{Fitzpatrick2015a,Fitzpatrick2014}, calculations of entanglement
entropy \cite{Hartman2013} and the bulk computation of classical
blocks from the geodesic approach
\cite{Hijano2015,Hijano2015a,Alkalaev:2015lca,Alkalaev:2016rjl,Chen:2016dfb}. In
fact, the isomonodromic approach generalizes the monodromy method of
\cite{Zamolodchikov1986,Hartman2013}. The main obstruction is solving
the PVI boundary condition in closed form. We gave a simple example
where the PVI solution simplifies and recovers the $c=\infty$ limit of
a linear dilaton conformal block \cite{Ribault:2014aa}. Many other
special limits remain to be explored, for example, the classical,
algebraic and Riccati solutions of PVI
\cite{Gamayun2012,Gamayun:2013auu}. Moreover, the connection between
the isomonodromic 
and the AGT approaches \cite{Ferrari2012a,Alba2011} can give further
insights on all these applications.

Although the relationship between classical conformal blocks, Painlevé
VI equation and isomonodromic deformations has been discussed before
\cite{Litvinov:2013sxa,Teschner2011,Teschner2017a}, the importance of
the $\tau$-function has been fully appreciated only here. This
relation is relevant not only from the technical point of view, but
also highlights the intriguing map between the $c=1$ and $c=\infty$
conformal blocks.  Coulomb Gas (Dotsenko--Fateev) integral
\cite{Mironov:2010zs} and Fredholm determinant
\cite{Bonelli:2016idi,Gavrylenko2016,Mironov:2017lgl} representations of conformal
blocks might also be useful to understand this map. 

Another important question is what is the quantum counterpart of the
$c=1$ isomonodromic structure described here. A related development is
that the canonical quantization of the isomonodromic equations is
equivalent to the 
Knizhnik--Zamolodchikov (KZ) equations (equivalently the isomonodromic
equations are the classical limit of the KZ
equations)~\cite{Reshetikhin1992,Harnad1994a}.  Solutions of BPZ
equations can be related to solutions of KZ equations
\cite{Sklyanin1989b,Stoyanovsky2000,Ribault2005}, connecting Liouville
theory to WZNW models. This web of relations, in connection to AGT
correspondence and Hitchin systems, was reviewed in
\cite{Teschner2011,Teschner2017a}. These papers also briefly mention
the role of the isomonodromic $\tau$-function in the semiclassical CFT
limit. More recently, the authors of \cite{Bershtein:2017swf} have
proposed a generalization of the isomonodromy/CFT correspondence to
arbitrary central charge using $q$-Painlevé conformal blocks
\cite{Bershtein:2016aef} and cluster algebras. It would be interesting
to study how the classical limit studied here emerges from this
quantum description.

The physical meaning of the $c=1$ and $c=\infty$ relationship, if any,
deserves further exploration in the future and we hope our work will
be helpful in this direction too.

\acknowledgments{
\label{sec:acknowledgments}

FN is grateful to Thomas Hartman, Oleg Lisovyy, Sergei Lukyanov and
Rubik Poghossian for important discussions on aspects of this work. FN
also thanks Bruno Carneiro da Cunha for his influential guidance on
the results presented here. FN and ML are supported by the Brazilian
ministries MCTI and MEC. The authors  acknowledge support by the
Simons Foundation, through the Simons Collaboration on the
Non-perturbative Bootstrap, during the Bootstrap School at ICTP-SAIFR,
São Paulo, where part of this work was done. }

\appendix

\section{Well-posedness of Initial Conditions}
\label{sec:well-posedn-lambd}

Here we show that the boundary conditions
\begin{equation}
  \label{eq:a1}
  \lambda(x) = x, \quad \mu(x) = -\frac{K_{x}}{2\theta_{t}}, 
\end{equation}
are well-posed with respect to the isomonodromic flow. The Garnier
system \eqref{eq:garniersystem} is given by 
\begin{subequations}
 \label{eq:garniersystemexplicit}
  \begin{align}
  \dot{\lambda} &= \frac{\lambda(\lambda-1)(\lambda-t)}{t(t-1)}
  \left(
    2\mu-\frac{2 \theta_0}{\lambda }-\frac{2 \theta_1}{\lambda -1}-\frac{2 \theta_t-1}{\lambda -t}
 \right),\\[5pt]
\dot{\mu} &= -\frac{1}{t(t-1)}
  \left[\,
(\lambda  (3 \lambda -2)+t(1-2 \lambda  ))\mu^{2} \right.\nonumber\\ &\left. +
(2 \theta _0 (-2 \lambda +t+1)+2 \theta _1 (t-2 \lambda )-(2 \lambda
            -1) \left(2 \theta _t-1\right))\mu + \kappa_{1}(\kappa_{2}+1)^{\phantom{1}}
  \right].
\end{align}
\end{subequations}
This system has three critical points at $t=0,1,\infty$. Close to these
singular points, the asymptotics of $\lambda(t)$ can be obtained via
Painlevé VI solutions
\cite{Guzzetti2012b}.  Our initial condition is defined close to a
regular point $t=x$, with $x\neq 0,1,\infty$. Therefore, the solution of
the Garnier system can be expressed as Taylor series in a neighborhood
around $t=x$. If we substitute the series solution around $t=x$
\begin{align}
  \label{eq:seriessolution}
  \lambda(t) &= x +
               \lambda_{1}(t-x)+\frac{\lambda_{2}}{2}(t-x)^{2}+\cdots,\\
\mu(t) &= \mu_{0} + \mu_{1}(t-x)+\cdots
\end{align}
into \eqref{eq:garniersystemexplicit}, we get that
$\lambda_{1}= 1-2\theta_{t}$ and that $\lambda_{2}$ and $\mu_{1}$ are
determined explicitly in terms of $x$, $\theta$'s and by
$\mu_{0}$. This means that $\mu_{0}$ can be taken as any finite
constant, for example, the initial condition \eqref{eq:a1}. Therefore,
the initial conditions \eqref{eq:a1} are consistent with a series
solution of the Garnier system around $t=x$.

\section{Ratio of Painlevé VI Structure Constants}
\label{sec:painleve-vi-tau}

Let $C_{n}\equiv C(\bm{\theta},\sigma+n)$, where $C(\btheta,\sigma)$
is given by \eqref{eq:structconsts}. A useful formula is the ratio of
two structure constants \cite{Gamayun2012}
   \begin{align}
     \label{eq:ratio}
     \frac{C_{n\pm 1}}{C_{n}}
  &=-\frac{\Gamma^2\left(1\mp2(\sigma+n)\right)}{\Gamma^2\left(1\pm2(\sigma+n)\right)}
    \prod_{\epsilon=\pm}\frac{\Gamma\left(1+\epsilon\theta_0+\theta_t\pm(\sigma+n)\right)
    \Gamma\left(1+\epsilon\theta_{\infty}+\theta_1\pm(\sigma+n)\right)}{
    \Gamma\left(1+\epsilon\theta_0+\theta_t\mp(\sigma+n)\right)\Gamma\left(1+\epsilon\theta_{\infty}+\theta_1\mp(\sigma+n)\right)}
    \times \nonumber\\
  & \times\,\frac{\left(\theta_0^2-(\theta_t\mp(\sigma+n))^2\right)
    \left(\theta_{\infty}^2-(\theta_1\mp(\sigma+n))^2\right)}{4\left(\sigma+n\right)^2\left(1\pm2(\sigma+n)\right)^2},
   \end{align}
   which can be derived from \eqref{eq:structconsts}. Using that
\begin{equation}
  \label{eq:7}
  \Gamma(z+n) = (z)_{n}\Gamma(z),\quad \Gamma(z-n)= \frac{(-1)^{n}\Gamma(z)}{(1-z)_{n}},
\end{equation}
where $(z)_{n}= z(z+1)\ldots(z+n-1)$ is the Pochhammer symbol, we have
\begin{multline}
  \label{eq:ratios}
  \frac{C_{n\pm1}}{C_{n}} = \left(
    \prod_{\epsilon=\pm}\frac{(1+\epsilon\theta_{0}+\theta_{t}+\sigma)_{n}
      (-\epsilon\theta_{0}-\theta_{t}+\sigma)_{n}
      (1+\epsilon\theta_{\infty}+\theta_{1}+\sigma)_{n}
      (-\epsilon\theta_{\infty}-\theta_{1}+\sigma)_{n}}{(2\sigma)_{2n\pm
        1}(1\pm 2\sigma)_{2n\pm1}} \right)^{\pm
    1}\\[5pt]\left(\theta_0^2-(\theta_t\mp(\sigma+n))^2\right)
  \left(\theta_{\infty}^2-(\theta_1\mp(\sigma+n))^2\right) 
\left(
    \frac{\theta_{\infty}+\theta_{1}+\sigma}{\theta_{\infty}+\theta_{1}-\sigma}
  \right)^{\pm 1}(-A)^{\pm 1}
\end{multline}
where
\begin{equation}
  \label{eq:10}
  [A(\btheta,\sigma)]^{\pm 1} \defeq
  \frac{4\sigma^{2}(1\pm 2\sigma)^{2}}{[(\theta_{\infty}\pm\sigma)^{2}-\theta_{1}^{2}][(\theta_{t}\mp\sigma)^{2}-\theta_{0}^{2}]}\frac{C_{\pm
      1}}{C_{0}}.
\end{equation}
Finally, we define the ratios
\begin{equation}
  \label{eq:fullratio}
  \bar{C}_{n} \equiv \frac{C_{n }}{C_{0}} =
 \prod_{k=0}^{|n|-1}\frac{C_{(k+1) \sgn(n)}}{C_{k\sgn(n)}} 
\end{equation}
and, if we use \eqref{eq:ratios}, we get
\begin{equation}
  \label{eq:fullratio2}
  \bar{C}_{n}(\btheta,\sigma) \defeq  \mathcal{C}_{n}(\btheta,\sgn(n)\sigma)A(\btheta,\sigma)^{n},
\end{equation}
where
\begin{multline}
  \label{eq:14}
  \mathcal{C}_{n}(\sigma) \equiv \\\prod_{k=0}^{|n|-1}
  \left(
    \prod_{\epsilon=\pm}\frac{(1+\epsilon\theta_{0}+\theta_{t}+\sigma)_{k}
      (-\epsilon\theta_{0}-\theta_{t}+\sigma)_{k}
      (1+\epsilon\theta_{\infty}+\theta_{1}+\sigma)_{k}
      (-\epsilon\theta_{\infty}-\theta_{1}+\sigma)_{k}}{(2\sigma)_{2k+
        1}(1+ 2\sigma)_{2k+1}} \right)\times
  \\[5pt]
  \times \left(
    \frac{\sigma+\theta_{\infty}+\theta_{1}}{\sigma-\theta_{\infty}-\theta_{1}}
  \right)
\left(\theta_0^2-(\theta_t-\sigma-k)^2\right)
  \left(\theta_{\infty}^2-(\theta_1-\sigma-k)^2\right).
\end{multline}


\section{Analytical and Numerical Checks of Accessory Parameter Expansion}
\label{sec:numer-checks-access}

As discussed in section \ref{sec:solving-lambdax=x}, the solution of
the condition $\lambda(x)=x$ gives a series expansion
$X(x) = 1+ X_{1}x + X_{2}x^{2}\ldots $ for the PVI integration
constant $s(x)$ encoded in $X(x)$. We present analytically the first
\begin{equation*}
  \label{eq:92}
    X_{1} = -\sigma \frac{ \left(\delta _{\sigma }^2+\left(\delta _1-\delta _{\infty }\right) \left(\delta _0-\delta _x\right)\right)}{\delta _{\sigma }^2},
\end{equation*}
and second order coefficients of this expansion
\begin{align*}
  \label{eq:Xcoeff2}
  X_{2} & =\sigma^{2}\frac{\left(\delta _{\sigma }^2+\left(\delta _1-\delta _{\infty }\right) \left(\delta _0-\delta _x\right)\right){}^2}{2 \delta _{\sigma }^4}+\nonumber\\[5pt]
        &+ \frac{\sigma}{8\delta_{\sigma}^{4}\left(4\delta_{\sigma}+3\right){}^{2}}
          \left\{
          \delta _1^2 \left[\delta _{\sigma }^2 \left(4 \delta _{\sigma } \left(\delta _{\sigma }+6\right)+9\right)+3 \left(20 \delta _{\sigma }^2-6 \delta _{\sigma }-9\right) \delta _x^2-6 \delta _{\sigma }^2 \left(8 \delta _{\sigma }+3\right) \delta _x\right]
          +\right. 
          \nonumber\\[5pt]
  &+\delta _{\sigma }^2 \left[-\delta _{\sigma }^2 \left(26 \delta
     _{\sigma } \left(2 \delta _{\sigma }+3\right)+27\right)+\left(4
     \delta _{\sigma } \left(\delta _{\sigma }+6\right)+9\right) \delta
     _x^2-6 \delta _{\sigma }^2 \delta _x\right] -
     \nonumber\\[5pt]
        &
          -6 \delta _{\sigma }^2 \delta _{\infty } \left[\delta _{\sigma
          }^2+\left(8 \delta _{\sigma }+3\right) \delta _x^2+2 \left(4 \delta
          _{\sigma } \left(\delta _{\sigma }+2\right)+3\right) \delta
          _x\right]+
          \nonumber\\[5pt]
        &+ \delta _0^2
          \left[6 \delta _1 \left[\left(-20 \delta _{\sigma }^2+6
          \delta _{\sigma }+9\right) \delta _{\infty }-\delta _{\sigma }^2
          \left(8 \delta _{\sigma }+3\right)\right]-6 \delta _{\sigma }^2
          \left(8 \delta _{\sigma }+3\right) \delta _{\infty
          }+\right.\nonumber\\[5pt]
        &\left.+3 \left(20
          \delta _{\sigma }^2-6 \delta _{\sigma }-9\right) \delta _{\infty
          }^2+3 \delta _1^2 \left(20 \delta _{\sigma }^2-6 \delta _{\sigma
          }-9\right)+\delta _{\sigma }^2 \left(4 \delta _{\sigma }
          \left(\delta _{\sigma }+6\right)+9\right)\right]+
          \nonumber\\[5pt]
        &+2 \delta _0 
          \left[6 \delta _1 \left[\delta _{\infty } \left[\left(8
          \delta _{\sigma }+3\right) \delta _{\sigma }^2+\left(20 \delta
          _{\sigma }^2-6 \delta _{\sigma }-9\right) \delta _x\right]+\delta
          _{\sigma }^2 \left(-4 \delta _{\sigma }^2+8 \delta _{\sigma }
          \left(\delta _x-1\right)+3 \left(\delta
          _x-1\right)\right)\right]-\right.\nonumber\\[5pt]
        &-3
          \delta _{\infty }^2 \left(\left(8 \delta _{\sigma }+3\right) \delta
          _{\sigma }^2+\left(20 \delta _{\sigma }^2-6 \delta _{\sigma
          }-9\right) \delta _x\right)+2 \delta _{\sigma }^2 \delta _{\infty }
          \left(4 \delta _{\sigma } \left(5 \delta _{\sigma }+6\right)+3
          \left(8 \delta _{\sigma }+3\right) \delta
          _x+9\right)-\nonumber\\[5pt]
        &\left.-3 \delta _1^2
          \left(\left(8 \delta _{\sigma }+3\right) \delta _{\sigma
          }^2+\left(20 \delta _{\sigma }^2-6 \delta _{\sigma }-9\right) \delta
          _x\right)+\delta _{\sigma }^2 \left(-3 \delta _{\sigma }^2-\left(4
          \delta _{\sigma } \left(\delta _{\sigma }+6\right)+9\right) \delta
          _x\right)\right]+
          \nonumber\\[5pt]
        &+2 \delta _1 \left(-\left(4 \delta _{\sigma } \left(\delta _{\sigma
          }+6\right)+9\right) \delta _{\sigma }^2 \delta _{\infty }-3 \delta
          _{\sigma }^4+2 \delta _{\sigma }^2 \delta _x \left(4 \delta _{\sigma
          } \left(5 \delta _{\sigma }+6 \delta _{\infty }+6\right)+9
          \left(\delta _{\infty }+1\right)\right)+\right.\nonumber\\[5pt]
        &\left.+3 \delta _x^2
          \left(\left(-20 \delta _{\sigma }^2+6 \delta _{\sigma }+9\right)
          \delta _{\infty }-\delta _{\sigma }^2 \left(8 \delta _{\sigma
          }+3\right)\right)\right)+
          \nonumber\\[5pt]
        &\left.+\delta _{\infty }^2 \left(+\left(\delta _{\sigma }^2 \left(4 \delta _{\sigma } \left(\delta _{\sigma }+6\right)+9\right)+3 \left(20 \delta _{\sigma }^2-6 \delta _{\sigma }-9\right) \delta _x^2-6 \delta _{\sigma }^2 \left(8 \delta _{\sigma }+3\right) \delta _x\right)\right)
          \right\}.
\end{align*}
Plugging back the $X(x)$ series into the logarithm of the
$\tau$-function \eqref{eq:normaltocanonical}, following the procedure
of section \ref{sec:access-param-expans}, we get the series expansion
of the accessory parameter
\begin{equation}
  \label{eq:88}
  H_{x} = \frac{\delta_{0}+\delta_{x}-\delta_{\sigma}}{x} +
  \frac{(\delta_{0}-\delta_{\sigma}-\delta_{x})(\delta_{\sigma}+\delta_{1}-\delta_{\infty})}{2\delta_{\sigma}}
 +\sum_{n=1}H_{n}x^{n}.
\end{equation}
We present here the next two terms in \eqref{eq:88}
\begin{dgroup*} 
\begin{dmath*}[style={\mathindent=0pt}]
    H_{1} = - \frac{x}{8 \delta _{\sigma }^3 \left(4 \delta
        _{\sigma}+3\right)} \left\{ \delta _0^2 \left(\delta _{\sigma
        }^2 \left(3-6 \delta _{\infty }\right)+5 \delta _{\sigma }
        \delta _{\infty }^2-2 \delta _1 \left(5 \delta _{\sigma }
          \delta _{\infty }+3 \delta _{\sigma }^2-3 \delta _{\infty
          }\right)+\delta _{\sigma }^3+\delta _1^2 \left(5 \delta
          _{\sigma }-3\right)-3 \delta _{\infty }^2\right) -2 \delta
      _0 \left(2 \delta _1 \left(3 \delta _{\sigma }^3-3 \delta
          _{\sigma }^2 \left(\delta _{\infty }+\delta _x-1\right)-5
          \delta _{\sigma } \delta _{\infty } \delta _x+3 \delta
          _{\infty } \delta _x\right)+\left(\delta _{\sigma }-\delta
          _{\infty }\right) \left(\delta _{\sigma }^2 \left(7 \delta
            _{\sigma }-3 \delta _{\infty }+6\right)+\delta _x
          \left(\delta _{\sigma } \left(3-5 \delta _{\infty
              }\right)+\delta _{\sigma }^2+3 \delta _{\infty
            }\right)\right)+\delta _1^2 \left(3 \delta _{\sigma
          }^2+\left(5 \delta _{\sigma }-3\right) \delta
          _x\right)\right) +\left(\delta _{\sigma }-\delta _{\infty
        }\right) \left(\delta _{\sigma }^2 \left(-\delta _{\sigma }
          \left(\delta _{\infty }-9\right)+13 \delta _{\sigma }^2-3
          \delta _{\infty }\right)+\delta _x^2 \left(\delta _{\sigma }
          \left(3-5 \delta _{\infty }\right)+\delta _{\sigma }^2+3
          \delta _{\infty }\right)+6 \delta _{\sigma }^2 \delta _x
        \left(3 \delta _{\sigma }+\delta _{\infty }+2\right)\right) -2
      \delta _1 \left(\delta _{\sigma }^2 \left(\delta _{\sigma }
          \left(\delta _{\infty }-6\right)-9 \delta _{\sigma }^2+3
          \delta _{\infty }\right)+\delta _x^2 \left(5 \delta _{\sigma
          } \delta _{\infty }+3 \delta _{\sigma }^2-3 \delta _{\infty
          }\right)-2 \delta _{\sigma }^2 \delta _x \left(5 \delta
          _{\sigma }+3 \delta _{\infty }+3\right)\right) +\delta _1^2
      \left(\delta _{\sigma }^2 \left(\delta _{\sigma
          }+3\right)+\left(5 \delta _{\sigma }-3\right) \delta _x^2-6
        \delta _{\sigma }^2 \delta _x\right) \right\},
 \end{dmath*}
 \begin{dmath*}
   H_{2} = \frac{1}{16 \delta _{\sigma }^5 \left(4 \delta _{\sigma
       }^2+11\delta _{\sigma }+6\right)} \left(\left(9 \delta _{\sigma
       }^2-19 \delta _{\sigma }+6\right) \delta _1^3+\left(-14 \delta
       _{\sigma }^3-3 \left(9 \delta _{\infty }-4\right) \delta
       _{\sigma }^2+57 \delta _{\infty } \delta _{\sigma }-18 \delta
       _{\infty }\right) \delta _1^2+\left(5 \delta _{\sigma }^4+7
       \delta _{\sigma }^3+3 \left(9 \delta _{\infty }^2-2\right)
       \delta _{\sigma }^2-57 \delta _{\infty }^2 \delta _{\sigma }+18
       \delta _{\infty }^2\right) \delta _1-\delta _{\infty } \left(5
       \delta _{\sigma }^4+\left(7-14 \delta _{\infty }\right) \delta
       _{\sigma }^3+3 \left(3 \delta _{\infty }^2+4 \delta _{\infty
         }-2\right) \delta _{\sigma }^2-19 \delta _{\infty }^2 \delta
       _{\sigma }+6 \delta _{\infty }^2\right)\right) \delta
   _0^3+\left(\left(2 \left(6-7 \delta _{\sigma }\right) \delta
       _{\sigma }^2-3 \delta _x \left(9 \delta _{\sigma }^2-19 \delta
         _{\sigma }+6\right)\right) \delta _1^3+\left(5 \delta
       _{\sigma }^4+21 \left(2 \delta _{\infty }+2 \delta _x-1\right)
       \delta _{\sigma }^3+9 \left(-4 \delta _{\infty }+\left(9 \delta
           _{\infty }-4\right) \delta _x+2\right) \delta _{\sigma
       }^2-171 \delta _x \delta _{\infty } \delta _{\sigma }+54 \delta
       _{\infty } \delta _x\right) \delta _1^2-3 \left(\delta _x
       \left(5 \delta _{\sigma }^4+7 \delta _{\sigma }^3+3 \left(9
           \delta _{\infty }^2-2\right) \delta _{\sigma }^2-57 \delta
         _{\infty }^2 \delta _{\sigma }+18 \delta _{\infty
         }^2\right)-2 \delta _{\sigma }^2 \left(2 \delta _{\sigma
         }^3+\left(5 \delta _{\infty }+4\right) \delta _{\sigma }^2-7
         \left(\delta _{\infty }-1\right) \delta _{\infty } \delta
         _{\sigma }+6 \left(\delta _{\infty }-1\right) \delta _{\infty
         }\right)\right) \delta _1-\left(\delta _{\sigma }-\delta
       _{\infty }\right) \left(3 \delta _{\sigma }^5-3 \left(7 \delta
         _{\infty }-5\right) \delta _{\sigma }^4+\left(14 \delta
         _{\infty }^2-3 \left(5 \delta _x+11\right) \delta _{\infty
         }+18\right) \delta _{\sigma }^3+3 \delta _{\infty } \left(-4
         \delta _{\infty }+\left(9 \delta _{\infty }-7\right) \delta
         _x+6\right) \delta _{\sigma }^2+3 \delta _{\infty }
       \left(6-19 \delta _{\infty }\right) \delta _x \delta _{\sigma
       }+18 \delta _{\infty }^2 \delta _x\right)\right) \delta
   _0^2+\left(\left(3 \left(9 \delta _{\sigma }^2-19 \delta _{\sigma
         }+6\right) \delta _x^2+\delta _{\sigma }^2 \left(5 \delta
         _{\sigma }^2+7 \delta _{\sigma }-6\right)\right) \delta
     _1^3-3 \left(\left(14 \delta _{\sigma }^3+3 \left(9 \delta
           _{\infty }-4\right) \delta _{\sigma }^2-57 \delta _{\infty
         } \delta _{\sigma }+18 \delta _{\infty }\right) \delta _x^2-2
       \delta _{\sigma }^2 \left(5 \delta _{\sigma }^2+7 \delta
         _{\sigma }-6\right) \delta _x-\delta _{\sigma }^2
       \left(\delta _{\sigma }+2\right) \left(4 \delta _{\sigma }^2-5
         \delta _{\infty } \delta _{\sigma }+3 \delta _{\infty
         }\right)\right) \delta _1^2+3 \left(\left(5 \delta _{\sigma
         }^4+7 \delta _{\sigma }^3+3 \left(9 \delta _{\infty
           }^2-2\right) \delta _{\sigma }^2-57 \delta _{\infty }^2
         \delta _{\sigma }+18 \delta _{\infty }^2\right) \delta _x^2-4
       \left(3 \delta _{\sigma }^2+5 \delta _{\infty } \delta _{\sigma
         }-3 \delta _{\infty }\right) \delta _{\sigma }^2 \left(\delta
         _{\sigma }+2\right) \delta _x+\left(7 \delta _{\sigma
         }^3+\left(9-12 \delta _{\infty }\right) \delta _{\sigma }^2+5
         \delta _{\infty }^2 \delta _{\sigma }-3 \delta _{\infty
         }^2\right) \delta _{\sigma }^2 \left(\delta _{\sigma
         }+2\right)\right) \delta _1+\left(\delta _{\sigma }-\delta
       _{\infty }\right) \left(3 \delta _{\infty } \left(-5 \delta
         _{\sigma }^3+\left(9 \delta _{\infty }-7\right) \delta
         _{\sigma }^2+\left(6-19 \delta _{\infty }\right) \delta
         _{\sigma }+6 \delta _{\infty }\right) \delta _x^2+6
       \left(\delta _{\sigma }^2+\left(3-5 \delta _{\infty }\right)
         \delta _{\sigma }+3 \delta _{\infty }\right) \delta _{\sigma
       }^2 \left(\delta _{\sigma }+2\right) \delta _x+\left(26 \delta
         _{\sigma }^3+\left(24-19 \delta _{\infty }\right) \delta
         _{\sigma }^2+\delta _{\infty } \left(5 \delta _{\infty
           }-3\right) \delta _{\sigma }-3 \delta _{\infty }^2\right)
       \delta _{\sigma }^2 \left(\delta _{\sigma
         }+2\right)\right)\right) \delta _0-\delta _1 \left(2 \left(19
       \delta _{\sigma }^2-3 \left(\delta _{\infty }-4\right) \delta
       _{\sigma }-9 \delta _{\infty }\right) \left(\delta _{\sigma
       }+2\right) \delta _{\sigma }^4-6 \delta _x^2 \left(4 \delta
       _{\sigma }^3+\left(5 \delta _{\infty }+8\right) \delta _{\sigma
       }^2+7 \delta _{\infty } \left(\delta _{\infty }+1\right) \delta
       _{\sigma }-6 \delta _{\infty } \left(\delta _{\infty
         }+1\right)\right) \delta _{\sigma }^2+3 \left(15 \delta
       _{\sigma }^3+3 \left(4 \delta _{\infty }+3\right) \delta
       _{\sigma }^2+5 \delta _{\infty }^2 \delta _{\sigma }-3 \delta
       _{\infty }^2\right) \delta _x \left(\delta _{\sigma }+2\right)
     \delta _{\sigma }^2+\left(5 \delta _{\sigma }^4+7 \delta _{\sigma
       }^3+3 \left(9 \delta _{\infty }^2-2\right) \delta _{\sigma
       }^2-57 \delta _{\infty }^2 \delta _{\sigma }+18 \delta _{\infty
       }^2\right) \delta _x^3\right)-\left(\delta _{\sigma }-\delta
     _{\infty }\right) \left(\left(23 \delta _{\sigma }^2-3
       \left(\delta _{\infty }-5\right) \delta _{\sigma }-9 \delta
       _{\infty }\right) \left(\delta _{\sigma }+2\right) \delta
     _{\sigma }^4+\left(3 \delta _{\sigma }^3+\left(15-9 \delta
         _{\infty }\right) \delta _{\sigma }^2+\left(-14 \delta
         _{\infty }^2-9 \delta _{\infty }+18\right) \delta _{\sigma
       }+6 \delta _{\infty } \left(2 \delta _{\infty }+3\right)\right)
     \delta _x^2 \delta _{\sigma }^2+\left(38 \delta _{\sigma
       }^3+\left(17 \delta _{\infty }+24\right) \delta _{\sigma
       }^2+\delta _{\infty } \left(5 \delta _{\infty }-3\right) \delta
       _{\sigma }-3 \delta _{\infty }^2\right) \delta _x \left(\delta
       _{\sigma }+2\right) \delta _{\sigma }^2+\delta _{\infty }
     \left(-5 \delta _{\sigma }^3+\left(9 \delta _{\infty }-7\right)
       \delta _{\sigma }^2+\left(6-19 \delta _{\infty }\right) \delta
       _{\sigma }+6 \delta _{\infty }\right) \delta _x^3\right)+\delta
   _1^2 \left(3 \left(\delta _{\sigma }^2+5 \delta _{\sigma }+6\right)
     \delta _{\sigma }^3-3 \delta _x \left(\delta _{\sigma }+2\right)
     \left(7 \delta _{\sigma }^2+\left(5 \delta _{\infty }-3\right)
       \delta _{\sigma }-3 \delta _{\infty }\right) \delta _{\sigma
     }+\left(14 \delta _{\sigma }^3+3 \left(9 \delta _{\infty
         }-4\right) \delta _{\sigma }^2-57 \delta _{\infty } \delta
       _{\sigma }+18 \delta _{\infty }\right) \delta _x^2\right)
   \left(\delta _x-\delta _{\sigma }\right)-\delta _1^3 \delta _x
   \left(\left(9 \delta _{\sigma }^2-19 \delta _{\sigma }+6\right)
     \delta _x^2-2 \delta _{\sigma }^2 \left(7 \delta _{\sigma
       }-6\right) \delta _x+\delta _{\sigma }^2 \left(5 \delta
       _{\sigma }^2+7 \delta _{\sigma }-6\right)\right).
\end{dmath*}
\end{dgroup*}
\begin{table}
\begin{center}
\begin{tabular}{|c|c|c|c|c|c|c|c|}
\hline 
{\footnotesize{}$\delta_{\sigma}$} & {\footnotesize{}$\delta_{0}$} & {\footnotesize{}$\delta_{1}$} & {\footnotesize{}$\delta_{x}$} & {\footnotesize{}$\delta_{\infty}$} & {\footnotesize{}$H_{3}$} & {\footnotesize{}$H_{4}$} & {\footnotesize{}$H_{5}$} 
\tabularnewline
\hline 
\hline 
{\footnotesize{}$1$} & {\footnotesize{}$1$} & {\footnotesize{}$1$} & {\footnotesize{}$1$} & {\footnotesize{}$1$} & {\footnotesize{}$-1.07867$} & {\footnotesize{}$-1.1431$1} & {\footnotesize{}$-1.18935$}
\tabularnewline
\hline 
{\footnotesize{}$1$} & {\footnotesize{}$0.4$} & {\footnotesize{}$0.4$} & {\footnotesize{}$0.4$} & {\footnotesize{}$0.4$} & {\footnotesize{}$-0.492431$} & {\footnotesize{}$-0.491792$} & {\footnotesize{}$-0.491431$} 
\tabularnewline
\hline 
{\footnotesize{}$2$} & {\footnotesize{}$1$} & {\footnotesize{}$1$} & {\footnotesize{}$2$} & {\footnotesize{}$2$} & {\footnotesize{}$-1.2865$} & {\footnotesize{}$-1.30949$} & {\footnotesize{}$-1.31137$} 
\tabularnewline
\hline 
{\footnotesize{}$2$} & {\footnotesize{}$2$} & {\footnotesize{}$2$} & {\footnotesize{}$1$} & {\footnotesize{}$1$} & {\footnotesize{}$-1.2865$} & {\footnotesize{}$-1.30949$} & {\footnotesize{}$-1.31137$} 
\tabularnewline
\hline 
{\footnotesize{}$0.5$} & {\footnotesize{}$0.1$} & {\footnotesize{}$0.2$} & {\footnotesize{}$0.3$} & {\footnotesize{}$0.4$} & {\footnotesize{}$-0.201122$} & {\footnotesize{}$-0.20042$} & {\footnotesize{}$-0.199985$} 
\tabularnewline
\hline 
{\footnotesize{}$0.5$} & {\footnotesize{}$0.2$} & {\footnotesize{}$0.1$} & {\footnotesize{}$0.4$} & {\footnotesize{}$0.3$} & {\footnotesize{}$-0.201122$} & {\footnotesize{}$-0.20042$} & {\footnotesize{}$-0.199985$} 
\tabularnewline
\hline 
{\footnotesize{}$0.4$} & {\footnotesize{}$2$} & {\footnotesize{}$1$} & {\footnotesize{}$2.1$} & {\footnotesize{}$1.2$} & {\footnotesize{}$-1.7517$} & {\footnotesize{}$-1.73116$} & {\footnotesize{}$-1.65458$} 
\tabularnewline
\hline 
{\footnotesize{}$0.4$} & {\footnotesize{}$2.1$} & {\footnotesize{}$1.2$} & {\footnotesize{}$2$} & {\footnotesize{}$1$} & {\footnotesize{}$-1.8517$} & {\footnotesize{}$-1.83116$} & {\footnotesize{}$-1.75458$} 
\tabularnewline
\hline 
    \end{tabular}\caption{Numerical coefficients from order $x^{3}$ to
      order $x^{5}$ for different values of $\delta$'s.}
\label{tab:numerics}
  \end{center}
\end{table}
We tested our results numerically from order $x^{3}$ and only up to
order $x^{5}$, as it is time consuming to simplify the expressions in
terms of $\delta$'s. The table \ref{tab:numerics} shows the values of
the coefficients $H_{n}$ if we substitute the $\theta$'s by numbers in
our algorithm from the beginning. The numbers presented here all match
the classical conformal block calculated via the inverse Gram matrix
CFT approach, also substituting numbers from the beginning.

Notice that we tested some special transformations of conformal blocks
by permutations of the values of the $\delta$'s, in the rows 3 to 8 of
the table. In the rows 3 and 4, and also 7 and 8, we permute
$\delta_{0}\leftrightarrow \delta_{x}, \delta_{1}\leftrightarrow
\delta_{\infty}$. The conformal block is not invariant under this
transformation, as we can see in rows 7 and 8. However, in the lines 3
and 4, the weights remain the same and this becomes a
symmetry. Finally, in rows 5 and 6, we test the symmetry
$\delta_{0}\leftrightarrow \delta_{1}, \delta_{x}\leftrightarrow
\delta_{\infty}$.

\bibliographystyle{JHEP} 
\bibliography{isoAGT}

\providecommand{\href}[2]{#2}\begingroup\raggedright\begin{thebibliography}{10}

\bibitem{Zamolodchikov1987}
A.~B. Zamolodchikov, {\it Conformal symmetry in two-dimensional space:
  recursion representation of conformal block},  {\em Theoretical and
  Mathematical Physics} {\bf 73} (1987), no.~1 1088--1093.

\bibitem{Zamolodchikov1996}
A.~B. Zamolodchikov and A.~B. Zamolodchikov, {\it {Structure constants and
  conformal bootstrap in Liouville field theory}},  {\em Nucl. Phys.} {\bf
  B477} (1996) 577--605, [\href{http://arxiv.org/abs/hep-th/9506136}{{\tt
  hep-th/9506136}}].

\bibitem{Litvinov:2013sxa}
A.~Litvinov, S.~Lukyanov, N.~Nekrasov, and A.~Zamolodchikov, {\it {Classical
  Conformal Blocks and Painleve VI}},  {\em JHEP} {\bf 07} (2014) 144,
  [\href{http://arxiv.org/abs/1309.4700}{{\tt arXiv:1309.4700}}].

\bibitem{Brown:1986nw}
J.~D. Brown and M.~Henneaux, {\it {Central Charges in the Canonical Realization
  of Asymptotic Symmetries: An Example from Three-Dimensional Gravity}},  {\em
  Commun. Math. Phys.} {\bf 104} (1986) 207--226.

\bibitem{Hartman2013}
T.~Hartman, {\it {Entanglement Entropy at Large Central Charge}},
  \href{http://arxiv.org/abs/1303.6955}{{\tt arXiv:1303.6955}}.

\bibitem{Fitzpatrick2014}
A.~L. Fitzpatrick, J.~Kaplan, and M.~T. Walters, {\it {Universality of
  Long-Distance AdS Physics from the CFT Bootstrap}},  {\em JHEP} {\bf 08}
  (2014) 145, [\href{http://arxiv.org/abs/1403.6829}{{\tt arXiv:1403.6829}}].

\bibitem{Hartman:2014oaa}
T.~Hartman, C.~A. Keller, and B.~Stoica, {\it {Universal Spectrum of 2d
  Conformal Field Theory in the Large c Limit}},  {\em JHEP} {\bf 09} (2014)
  118, [\href{http://arxiv.org/abs/1405.5137}{{\tt arXiv:1405.5137}}].

\bibitem{Fitzpatrick2015a}
A.~L. Fitzpatrick, J.~Kaplan, and M.~T. Walters, {\it {Virasoro Conformal
  Blocks and Thermality from Classical Background Fields}},  {\em JHEP} {\bf
  11} (2015) 200, [\href{http://arxiv.org/abs/1501.05315}{{\tt
  arXiv:1501.05315}}].

\bibitem{Ryu:2006bv}
S.~Ryu and T.~Takayanagi, {\it {Holographic derivation of entanglement entropy
  from AdS/CFT}},  {\em Phys. Rev. Lett.} {\bf 96} (2006) 181602,
  [\href{http://arxiv.org/abs/hep-th/0603001}{{\tt hep-th/0603001}}].

\bibitem{Hijano2015}
E.~Hijano, P.~Kraus, E.~Perlmutter, and R.~Snively, {\it {Semiclassical
  Virasoro blocks from AdS$_{3}$ gravity}},  {\em JHEP} {\bf 12} (2015) 077,
  [\href{http://arxiv.org/abs/1508.04987}{{\tt arXiv:1508.04987}}].

\bibitem{Hijano2015a}
E.~Hijano, P.~Kraus, and R.~Snively, {\it {Worldline approach to semi-classical
  conformal blocks}},  {\em JHEP} {\bf 07} (2015) 131,
  [\href{http://arxiv.org/abs/1501.02260}{{\tt arXiv:1501.02260}}].

\bibitem{Alkalaev:2015lca}
K.~B. Alkalaev and V.~A. Belavin, {\it {Monodromic vs geodesic computation of
  Virasoro classical conformal blocks}},  {\em Nucl. Phys.} {\bf B904} (2016)
  367--385, [\href{http://arxiv.org/abs/1510.06685}{{\tt arXiv:1510.06685}}].

\bibitem{Alkalaev:2016rjl}
K.~B. Alkalaev, {\it {Many-point classical conformal blocks and geodesic
  networks on the hyperbolic plane}},  {\em JHEP} {\bf 12} (2016) 070,
  [\href{http://arxiv.org/abs/1610.06717}{{\tt arXiv:1610.06717}}].

\bibitem{Chen:2016dfb}
B.~Chen, J.-q. Wu, and J.-j. Zhang, {\it {Holographic Description of 2D
  Conformal Block in Semi-classical Limit}},  {\em JHEP} {\bf 10} (2016) 110,
  [\href{http://arxiv.org/abs/1609.00801}{{\tt arXiv:1609.00801}}].

\bibitem{Roberts:2014ifa}
D.~A. Roberts and D.~Stanford, {\it {Two-dimensional conformal field theory and
  the butterfly effect}},  {\em Phys. Rev. Lett.} {\bf 115} (2015), no.~13
  131603, [\href{http://arxiv.org/abs/1412.5123}{{\tt arXiv:1412.5123}}].

\bibitem{Harlow:2011ny}
D.~Harlow, J.~Maltz, and E.~Witten, {\it {Analytic Continuation of Liouville
  Theory}},  {\em JHEP} {\bf 12} (2011) 071,
  [\href{http://arxiv.org/abs/1108.4417}{{\tt arXiv:1108.4417}}].

\bibitem{Dong:2018esp}
X.~Dong, S.~Maguire, A.~Maloney, and H.~Maxfield, {\it {Phase transitions in 3D
  gravity and fractal dimension}},  \href{http://arxiv.org/abs/1802.07275}{{\tt
  arXiv:1802.07275}}.

\bibitem{Belavin1984}
A.~A. Belavin, A.~M. Polyakov, and A.~B. Zamolodchikov, {\it {Infinite
  Conformal Symmetry in Two-Dimensional Quantum Field Theory}},  {\em Nucl.
  Phys.} {\bf B241} (1984) 333--380.

\bibitem{Zamolodchikov1984}
A.~B. Zamolodchikov, {\it {Conformal Symmetry In Two-Dimensions: An Explicit
  Recurrence Formula For The Conformal Partial Wave Amplitude}},  {\em Commun.
  Math. Phys.} {\bf 96} (1984) 419--422.

\bibitem{Perlmutter2015}
E.~Perlmutter, {\it {Virasoro Conformal Blocks in Closed Form}},  {\em JHEP}
  {\bf 08} (2015) 088, [\href{http://arxiv.org/abs/1502.07742}{{\tt
  arXiv:1502.07742}}].

\bibitem{Alday2010b}
L.~F. Alday, D.~Gaiotto, and Y.~Tachikawa, {\it {Liouville Correlation
  Functions from Four-dimensional Gauge Theories}},  {\em Lett. Math. Phys.}
  {\bf 91} (2010) 167--197, [\href{http://arxiv.org/abs/0906.3219}{{\tt
  arXiv:0906.3219}}].

\bibitem{Nekrasov2003a}
N.~A. Nekrasov, {\it {Seiberg-Witten Prepotential From Instanton Counting}},
  {\em Adv. Theor. Math. Phys.} {\bf 7} (2003), no.~5 831--864,
  [\href{http://arxiv.org/abs/hep-th/0206161}{{\tt hep-th/0206161}}].

\bibitem{Nekrasov2009}
N.~A. Nekrasov and S.~L. Shatashvili, {\it {Quantization of Integrable Systems
  and Four Dimensional Gauge Theories}},  in {\em {Proceedings, 16th
  International Congress on Mathematical Physics (ICMP09): Prague, Czech
  Republic, August 3-8, 2009}}, pp.~265--289, 2009.
\newblock \href{http://arxiv.org/abs/0908.4052}{{\tt arXiv:0908.4052}}.

\bibitem{Piatek2011}
M.~Piatek, {\it {Classical conformal blocks from TBA for the elliptic
  Calogero-Moser system}},  {\em JHEP} {\bf 06} (2011) 050,
  [\href{http://arxiv.org/abs/1102.5403}{{\tt arXiv:1102.5403}}].

\bibitem{Ferrari2012a}
F.~Ferrari and M.~Piatek, {\it {Liouville theory, N=2 gauge theories and
  accessory parameters}},  {\em JHEP} {\bf 05} (2012) 025,
  [\href{http://arxiv.org/abs/1202.2149}{{\tt arXiv:1202.2149}}].

\bibitem{Nekrasov:2011bc}
N.~Nekrasov, A.~Rosly, and S.~Shatashvili, {\it {Darboux coordinates, Yang-Yang
  functional, and gauge theory}},  {\em Nucl.Phys.Proc.Suppl.} {\bf 216} (Mar.,
  2011) 69--93, [\href{http://arxiv.org/abs/1103.3919}{{\tt arXiv:1103.3919}}].

\bibitem{Teschner2016}
J.~Teschner, {\it {Exact Results on $\mathcal{N} =$ 2 Supersymmetric Gauge
  Theories}},  in {\em New Dualities of Supersymmetric Gauge Theories}
  (J.~Teschner, ed.), pp.~1--30.
\newblock 2016.
\newblock \href{http://arxiv.org/abs/1412.7145}{{\tt arXiv:1412.7145}}.

\bibitem{Guzzetti2012b}
D.~Guzzetti, {\it Tabulation of painlev{\'e} 6 transcendents},  {\em
  Nonlinearity} {\bf 25} (2012), no.~12 3235.

\bibitem{ronveaux1995heun}
A.~Ronveaux and F.~Arscott, {\em Heun's differential equations}.
\newblock Oxford University Press, 1995.

\bibitem{Jimbo1981b}
M.~Jimbo, T.~Miwa, and A.~K. Ueno, {\it {Monodromy Preserving Deformation of
  Linear Ordinary Differential Equations With Rational Coefficients, I}},  {\em
  Physica} {\bf D2} (1981) 306--352.

\bibitem{Jimbo:1981-2}
M.~Jimbo and T.~Miwa, {\it {Monodromy Preserving Deformation of Linear Ordinary
  Differential Equations with Rational Coefficients, II}},  {\em Physica} {\bf
  D2} (1981) 407--448.

\bibitem{Jimbo:1981-3}
M.~Jimbo and T.~Miwa, {\it {Monodromy Preserving Deformation of Linear Ordinary
  Differential Equations with Rational Coefficients, III}},  {\em Physica} {\bf
  D4} (1981) 26--46.

\bibitem{Jimbo:1982}
M.~Jimbo, {\it {Monodromy Problem and the boundary condition for some
  Painlev{\'e} equations}},  {\em Publ. Res. Inst. Math. Sci.} {\bf 18} (1982)
  1137--1161.

\bibitem{Gamayun2012}
O.~Gamayun, N.~Iorgov, and O.~Lisovyy, {\it {Conformal field theory of
  Painlev\'e VI}},  {\em JHEP} {\bf 10} (2012) 038,
  [\href{http://arxiv.org/abs/1207.0787}{{\tt arXiv:1207.0787}}]. [Erratum:
  JHEP10,183(2012)].

\bibitem{Sato1978}
M.~Sato, T.~Miwa, and M.~Jimbo, {\it {Holonomic quantum fields I}},  {\em
  Publications of the Research Institute for Mathematical Sciences} {\bf 14}
  (1978), no.~1 223--267.

\bibitem{Sato1979}
M.~Sato, T.~Miwa, and M.~Jimbo, {\it {Holonomic Quantum Fields. II---The
  Riemann-Hilbert Problem---}},  {\em Publications of the Research Institute
  for Mathematical Sciences} {\bf 15} (1979), no.~1 201--278.

\bibitem{Sato1979b}
M.~Sato, T.~Miwa, and M.~Jimbo, {\it {Holonomic quantum fields III}},  {\em
  Publications of the Research Institute for Mathematical Sciences} {\bf 15}
  (1979), no.~2 577--629.

\bibitem{Sato1979a}
M.~Sato, T.~Miwa, and M.~Jimbo, {\it {Holonomic quantum fields. IV}},  {\em
  Publications of the Research Institute for Mathematical Sciences} {\bf 15}
  (1979), no.~3 871--972.

\bibitem{Sato1980}
M.~Sato, T.~Miwa, and M.~Jimbo, {\it {Holonomic quantum fields. V}},  {\em
  Publications of the Research Institute for Mathematical Sciences} {\bf 16}
  (1980), no.~2 531--584.

\bibitem{Iorgov2015}
N.~Iorgov, O.~Lisovyy, and J.~Teschner, {\it {Isomonodromic tau-functions from
  Liouville conformal blocks}},  {\em Commun. Math. Phys.} {\bf 336} (2015),
  no.~2 671--694, [\href{http://arxiv.org/abs/1401.6104}{{\tt
  arXiv:1401.6104}}].

\bibitem{Teschner2011}
J.~Teschner, {\it {Quantization of the Hitchin moduli spaces, Liouville theory,
  and the geometric Langlands correspondence I}},  {\em Adv. Theor. Math.
  Phys.} {\bf 15} (2011), no.~2 471--564,
  [\href{http://arxiv.org/abs/1005.2846}{{\tt arXiv:1005.2846}}].

\bibitem{Teschner2017a}
J.~Teschner, {\it {Classical conformal blocks and isomonodromic deformations}},
   \href{http://arxiv.org/abs/1707.07968}{{\tt arXiv:1707.07968}}.

\bibitem{Gamayun:2013auu}
O.~Gamayun, N.~Iorgov, and O.~Lisovyy, {\it {How instanton combinatorics solves
  Painlev{\'e} VI, V and IIIs}},  {\em J.Phys.} {\bf A46} (Feb., 2013) 335203,
  [\href{http://arxiv.org/abs/1302.1832}{{\tt arXiv:1302.1832}}].

\bibitem{Chang2016}
C.-M. Chang and Y.-H. Lin, {\it {Bootstrapping 2D CFTs in the Semiclassical
  Limit}},  {\em JHEP} {\bf 08} (2016) 056,
  [\href{http://arxiv.org/abs/1510.02464}{{\tt arXiv:1510.02464}}].

\bibitem{Chang2016a}
C.-M. Chang and Y.-H. Lin, {\it {Bootstrap, universality and horizons}},  {\em
  JHEP} {\bf 10} (2016) 068, [\href{http://arxiv.org/abs/1604.01774}{{\tt
  arXiv:1604.01774}}].

\bibitem{Collier2016}
S.~Collier, Y.-H. Lin, and X.~Yin, {\it {Modular Bootstrap Revisited}},
  \href{http://arxiv.org/abs/1608.06241}{{\tt arXiv:1608.06241}}.

\bibitem{Collier2017}
S.~Collier, P.~Kravchuk, Y.-H. Lin, and X.~Yin, {\it {Bootstrapping the
  Spectral Function: On the Uniqueness of Liouville and the Universality of
  BTZ}},  \href{http://arxiv.org/abs/1702.00423}{{\tt arXiv:1702.00423}}.

\bibitem{Ribault:2014aa}
S.~Ribault, {\it {Conformal field theory on the plane}},
  \href{http://arxiv.org/abs/1406.4290}{{\tt arXiv:1406.4290}}.

\bibitem{Teschner2017}
J.~Teschner, {\it {A guide to two-dimensional conformal field theory}},
  \href{http://arxiv.org/abs/1708.00680}{{\tt arXiv:1708.00680}}.

\bibitem{Iwasaki:1991}
K.~Iwasaki, H.~Kimura, S.~Shimomura, and M.~Yoshida, {\em {From Gauss to
  Painlev{\'e}: A Modern Theory of Special Functions}}, vol.~16 of {\em Aspects
  of Mathematics E}.
\newblock Braunschweig, 1991.

\bibitem{Hollands:2017ahy}
L.~Hollands and O.~Kidwai, {\it {Higher length-twist coordinates, generalized
  Heun's opers, and twisted superpotentials}},
  \href{http://arxiv.org/abs/1710.04438}{{\tt arXiv:1710.04438}}.

\bibitem{Schlesinger1912}
L.~Schlesinger, {\it {\"U}ber eine klasse von differentialsystemen beliebiger
  ordnung mit festen kritischen punkten.},  {\em Journal f{\"u}r die reine und
  angewandte Mathematik} {\bf 141} (1912) 96--145.

\bibitem{Garnier:1912}
R.~Garnier, {\it {Sur des {\'e}quations diff{\'e}rentielles du troisi{\`e}me
  ordre dont l'int{\'e}grale g{\'e}n{\'e}rale est uniforme et sur une classe
  d'{\'e}quations nouvelles d'ordre sup{\'e}rieur dont l'int{\'e}grale
  g{\'e}n{\'e}rale {\`a} ses points critiques fixes}},  {\em Ann. Ecol. Norm.
  Sup.} {\bf 29} (1912) 1.

\bibitem{Novikov2009}
D.~P. Novikov, {\it {The 2$\times$ 2 matrix Schlesinger system and the
  Belavin-Polyakov-Zamolodchikov system}},  {\em Theoretical and Mathematical
  Physics} {\bf 161} (2009), no.~2 1485--1496.

\bibitem{Novaes2014c}
F.~Novaes and B.~Carneiro~da Cunha, {\it {Isomonodromy, Painlev{\'e}
  transcendents and scattering off of black holes}},  {\em JHEP} {\bf 07}
  (2014) 132, [\href{http://arxiv.org/abs/1404.5188}{{\tt arXiv:1404.5188}}].

\bibitem{daCunha:2015fna}
B.~Carneiro~da Cunha and F.~Novaes, {\it {Kerr--de Sitter greybody factors via
  isomonodromy}},  {\em Phys. Rev.} {\bf D93} (2016), no.~2 024045,
  [\href{http://arxiv.org/abs/1508.04046}{{\tt arXiv:1508.04046}}].

\bibitem{Bershtein:2014yia}
M.~A. Bershtein and A.~I. Shchechkin, {\it {Bilinear equations on Painlev{\'e}
  $\tau$ functions from CFT}},  {\em Commun. Math. Phys.} {\bf 339} (2015),
  no.~3 1021--1061, [\href{http://arxiv.org/abs/1406.3008}{{\tt
  arXiv:1406.3008}}].

\bibitem{Okamoto1986}
K.~Okamoto, {\it {Studies on the Painlev{\'e} equations}},  {\em Annali di
  Matematica pura ed applicata} {\bf 146} (1986), no.~1 337--381.

\bibitem{Mironov:2010zs}
A.~Mironov, A.~Morozov, and S.~Shakirov, {\it {Conformal blocks as
  Dotsenko-Fateev Integral Discriminants}},  {\em Int. J. Mod. Phys.} {\bf A25}
  (2010) 3173--3207, [\href{http://arxiv.org/abs/1001.0563}{{\tt
  arXiv:1001.0563}}].

\bibitem{Witten1988}
E.~Witten, {\it {Coadjoint Orbits of the Virasoro Group}},  {\em Commun. Math.
  Phys.} {\bf 114} (1988) 1.

\bibitem{Raeymaekers2015}
J.~Raeymaekers, {\it {Quantization of conical spaces in 3D gravity}},  {\em
  JHEP} {\bf 03} (2015) 060, [\href{http://arxiv.org/abs/1412.0278}{{\tt
  arXiv:1412.0278}}].

\bibitem{Hulik2017}
O.~Hul{\'\i}k, T.~Proch{\'a}zka, and J.~Raeymaekers, {\it {Multi-centered
  AdS$_{3}$ solutions from Virasoro conformal blocks}},  {\em JHEP} {\bf 03}
  (2017) 129, [\href{http://arxiv.org/abs/1612.03879}{{\tt arXiv:1612.03879}}].

\bibitem{iorgov2013painleve}
N.~Iorgov, O.~Lisovyy, and Y.~Tykhyy, {\it {Painlev{\'e} VI connection problem
  and monodromy of c= 1 conformal blocks}},  {\em Journal of High Energy
  Physics} {\bf 2013} (2013), no.~12 29.

\bibitem{Piatek2017}
M.~Piatek and A.~R. Pietrykowski, {\it {Solving Heun's equation using conformal
  blocks}},  \href{http://arxiv.org/abs/1708.06135}{{\tt arXiv:1708.06135}}.

\bibitem{Amado:2017kao}
J.~B. Amado, B.~Carneiro~da Cunha, and E.~Pallante, {\it {On the Kerr-AdS/CFT
  correspondence}},  {\em JHEP} {\bf 08} (2017) 094,
  [\href{http://arxiv.org/abs/1702.01016}{{\tt arXiv:1702.01016}}].

\bibitem{Bonelli2016}
G.~Bonelli, O.~Lisovyy, K.~Maruyoshi, A.~Sciarappa, and A.~Tanzini, {\it {On
  Painlev\'e/gauge theory correspondence}},
  \href{http://arxiv.org/abs/1612.06235}{{\tt arXiv:1612.06235}}.

\bibitem{Gaiotto2013}
D.~Gaiotto, {\it {Asymptotically free $\mathcal{N} = 2$ theories and irregular
  conformal blocks}},  {\em J. Phys. Conf. Ser.} {\bf 462} (2013), no.~1
  012014, [\href{http://arxiv.org/abs/0908.0307}{{\tt arXiv:0908.0307}}].

\bibitem{Piatek2014}
M.~Piatek and A.~R. Pietrykowski, {\it {Classical irregular block, $
  \mathcal{N} $ = 2 pure gauge theory and Mathieu equation}},  {\em JHEP} {\bf
  12} (2014) 032, [\href{http://arxiv.org/abs/1407.0305}{{\tt
  arXiv:1407.0305}}].

\bibitem{Nagoya2015}
H.~Nagoya, {\it {Irregular conformal blocks, with an application to the fifth
  and fourth Painlev{\'e} equations}},  {\em J. Math. Phys.} {\bf 56} (2015),
  no.~12 123505, [\href{http://arxiv.org/abs/1505.02398}{{\tt
  arXiv:1505.02398}}].

\bibitem{Gavrylenko2016}
P.~Gavrylenko and O.~Lisovyy, {\it {Fredholm determinant and Nekrasov sum
  representations of isomonodromic tau functions}},
  \href{http://arxiv.org/abs/1608.00958}{{\tt arXiv:1608.00958}}.

\bibitem{Zamolodchikov1986}
A.~B. Zamolodchikov, {\it {Two-dimensional conformal symmetry and critical
  four-spin correlation functions in the Ashkin--Teller model}},  {\em Sov.
  Phys.-JETP} {\bf 63} (1986) 1061--1066.

\bibitem{Alba2011}
V.~A. Alba, V.~A. Fateev, A.~V. Litvinov, and G.~M. Tarnopolskiy, {\it {On
  combinatorial expansion of the conformal blocks arising from AGT
  conjecture}},  {\em Lett. Math. Phys.} {\bf 98} (2011) 33--64,
  [\href{http://arxiv.org/abs/1012.1312}{{\tt arXiv:1012.1312}}].

\bibitem{Bonelli:2016idi}
G.~Bonelli, A.~Grassi, and A.~Tanzini, {\it {Seiberg--Witten theory as a Fermi
  gas}},  {\em Lett. Math. Phys.} {\bf 107} (2017), no.~1 1--30,
  [\href{http://arxiv.org/abs/1603.01174}{{\tt arXiv:1603.01174}}].

\bibitem{Mironov:2017lgl}
A.~Mironov and A.~Morozov, {\it {On determinant representation and
  integrability of Nekrasov functions}},  {\em Phys. Lett.} {\bf B773} (2017)
  34--46, [\href{http://arxiv.org/abs/1707.02443}{{\tt arXiv:1707.02443}}].

\bibitem{Reshetikhin1992}
N.~Reshetikhin, {\it {The Knizhnik-Zamolodchikov system as a deformation of the
  isomonodromy problem}},  {\em letters in mathematical physics} {\bf 26}
  (1992), no.~3 167--177.

\bibitem{Harnad1994a}
J.~P. Harnad, {\it {Quantum isomonodromic deformations and the
  Knizhnik-Zamolodchikov equations}},  in {\em {Symmetries and Integrability of
  Difference Equations}}, pp.~155--161, 1994.
\newblock \href{http://arxiv.org/abs/hep-th/9406078}{{\tt hep-th/9406078}}.

\bibitem{Sklyanin1989b}
E.~K. Sklyanin, {\it {Separation of variables in the Gaudin model}},  {\em J.
  Sov. Math.} {\bf 47} (1989) 2473--2488. [Zap. Nauchn. Semin.164,151(1987)].

\bibitem{Stoyanovsky2000}
A.~V. Stoyanovsky, {\it {A relation between the Knizhnik-Zamolodchikov and
  Belavin-Polyakov-Zamolodchikov systems of partial differential equations}},
  \href{http://arxiv.org/abs/math-ph/0012013}{{\tt math-ph/0012013}}.

\bibitem{Ribault2005}
S.~Ribault and J.~Teschner, {\it {H+(3)-WZNW correlators from Liouville
  theory}},  {\em JHEP} {\bf 06} (2005) 014,
  [\href{http://arxiv.org/abs/hep-th/0502048}{{\tt hep-th/0502048}}].

\bibitem{Bershtein:2017swf}
M.~Bershtein, P.~Gavrylenko, and A.~Marshakov, {\it {Cluster integrable
  systems, $q$-Painlev{\'e} equations and their quantization}},  {\em JHEP}
  {\bf 02} (2018) 077, [\href{http://arxiv.org/abs/1711.02063}{{\tt
  arXiv:1711.02063}}].

\bibitem{Bershtein:2016aef}
M.~A. Bershtein and A.~I. Shchechkin, {\it {q-deformed Painlev{\'e} $\tau$
  function and q-deformed conformal blocks}},  {\em J. Phys.} {\bf A50} (2017),
  no.~8 085202, [\href{http://arxiv.org/abs/1608.02566}{{\tt
  arXiv:1608.02566}}].

\end{thebibliography}\endgroup

\end{document}